\documentclass[a4paper, 11pt]{article}
\usepackage{latexsym,amsmath,amsfonts,amssymb}
\usepackage{tikz}
\usetikzlibrary{decorations.pathmorphing,cd,decorations.markings,calc}
\usepackage{mathrsfs}
\usepackage[american]{babel}
\usepackage{graphicx}
\usepackage{bbm}
\usepackage{cite}
\usepackage{tcolorbox}

\usepackage[colorlinks=true, citecolor=blue!90!black, linkcolor=blue!90!black, linktocpage=true, urlcolor=red!70!black]{hyperref}

\renewcommand{\baselinestretch}{1.2}
\newcommand{\JAD}[1]{{\bf [{JAD:} #1]}}

\newcommand{\ie}{\textit{i.e.}}

\numberwithin{equation}{section}

\newcommand{\be}{\begin{equation}} \newcommand{\ee}{\end{equation}}
\newcommand{\bea}{\begin{equation} \begin{aligned}} \newcommand{\eea}{\end{aligned} \end{equation}}

\newcommand{\cL}{\mathcal{L}}

\newcommand{\cM}{\mathcal{M}}

\newcommand{\cN}{\mathcal{N}}
\newcommand{\cO}{\mathcal{O}}

\newcommand{\cW}{\mathcal{W}}

\newcommand{\bC}{\mathbb{C}}

\newcommand{\bP}{\mathbb{P}}

\newcommand{\bR}{\mathbb{R}}

\newcommand{\bZ}{\mathbb{Z}}

\newcommand{\unit}{\mathbbm{1}}

\def\repa{\raise4pt\hbox{$\square$}\mkern-14mu\raise-4pt\hbox{$\square$}}
\def\repab{\overline{\raise4pt\hbox{$\square$}\mkern-14mu\raise-4pt\hbox{$\square$}\mkern-1mu}}

\DeclareMathOperator{\Tr}{Tr}

\begin{document}

\thispagestyle{empty}
\fontsize{12pt}{20pt}
\vspace{13mm}
\begin{center}
	{\huge 
    Confining Strings in a Gapless Phase 
	}\\[13mm]
    	{\large Jeremias Aguilera Damia$^{a}$, Giovanni Galati$^{b}$, and Giovanni Rizi$^{c}$}
	
	\bigskip
	{\it  $^a$ Departament de F\' \i sica Cu\' antica i Astrof\'\i sica and Institut de Ci\`encies del Cosmos,\\ 
Universitat de Barcelona, Mart\'i Franqu\`es, 1, ES-08028
Barcelona, Spain \\ $^b$ Physique Th\'eorique et Math\'ematique and International Solvay Institutes\\
Universit\'e Libre de Bruxelles, C.P. 231, 1050 Brussels, Belgium  \\
		$^c$ Institut des Hautes Etudes Scientifiques, 91440 Bures-sur-Yvette, France
        \\
      
	}
\end{center}

\bigskip

\begin{abstract}

\noindent We consider the dynamics of confined strings embedded in a gapless four-dimensional theory. To this end, we examine finite-tension string-like solutions to the equations of motion of the $\bC\bP^1$ non-linear sigma model. We present a comprehensive analysis of the quantum fluctuations around these solutions and derive the corresponding spectrum. These results allow us to determine the quantum corrections to the closed string ground state energy in both the finite- and infinite-size limits. Furthermore, we analyze quantum corrections to the string's effective width. We find that these observables generically depart from the universal predictions of standard Effective String Theory (EST), and we identify specific limits in which the bulk dynamics decouple and EST is recovered. Finally, we discuss the connection between these string configurations and stable electric and magnetic fluxes arising in certain ultraviolet completions of the $\bC\bP^1$ model.

\end{abstract}

\newpage
\pagenumbering{arabic}
\setcounter{page}{1}
\setcounter{footnote}{0}
\renewcommand{\thefootnote}{\arabic{footnote}}

{\renewcommand{\baselinestretch}{.88} \parskip=0pt
\setcounter{tocdepth}{2}
\tableofcontents}

\newpage

\section{Introduction}

Extended excitations arise naturally in a broad class of quantum field theories. Among these, string--like objects play a prominent role, with notable examples including flux tubes in confining gauge theories and localized magnetic flux configurations in four--dimensional superconductors. In non--Abelian gauge theories such as Yang--Mills or QCD, the emergence of such strings is closely connected to the mechanism of color confinement, one of the most fundamental yet still poorly understood features of strong interactions. Below the confinement scale $\Lambda$, the energy carried by the gauge fields sourced by color charges is squeezed into thin tubular regions, giving rise to flux tubes characterized by a finite tension. The dynamics of such string-like excitations often captures valuable physical information about the underlying theory, including the spectrum and interactions of mesonic and baryonic states in the confined phase. A defining property of these objects is their tension $T$, which corresponds to a constant vacuum energy density on the string worldsheet and is generically set by the characteristic scale $\Lambda$ governing the effective bulk description. However, in most physically relevant situations, a precise identification of the low-energy degrees of freedom localized on the string worldsheet, together with a controlled derivation of their dynamics from first principles, remains elusive. The main obstacle is that the bulk theory typically becomes strongly coupled at the scale $\Lambda$, as is the case for confining flux tubes in asymptotically free gauge theories. This strongly coupled nature of the bulk usually limits top-down analysis of confining strings.

In typical confining scenarios, the onset of confinement is accompanied by the generation of a mass gap $M_{\rm gap} \sim \Lambda$, and the low-energy spectrum above the gapped vacuum is expected to be captured by a universal two-dimensional Effective String Theory (EST) describing the dynamics of long flux tubes and their lightest excitations \cite{NAMBU1979372,POLYAKOV1980171,Luscher:1980iy,LUSCHER1981317,Polchinski:1991ax}. In these cases, for a given string of length $L$ embedded in $D>2$ spacetime dimensions, there is a universal sector of light worldsheet degrees of freedom accounted for the $(D-2)$ massless Nambu--Goldstone (NG) fields associated to the spacetime symmetry breaking pattern
 \be
 ISO(1,D-1)\to ISO(1,1)\times O(D-2)\,.
 \ee
(see for instance \cite{Low:2001bw} for a discussion on the counting of NG bosons). Interactions among the Goldstone modes are suppressed at low worldsheet momenta $q \sim L^{-1}$. Consequently, this universal sector admits a systematic description of long string states in the regime $L^2 \gg T^{-1}$, organized as a derivative expansion in which higher–order contributions are suppressed by inverse powers of $T L^2$. 
This framework has by now become a well–established and actively investigated subject, see e.g. \cite{Aharony:2011gb,Aharony:2009gg,Dubovsky:2012sh,Aharony:2013ipa,Dubovsky:2015zey,Brandt:2016xsp,EliasMiro:2019kyf,Komargodski:2024swh,Cuomo:2024gek,Gabai:2025hwf} for a non–exhaustive list of references.\footnote{The case $D=2$ is special since flux tubes are Lorentz invariant and there are no corresponding gapless excitations. The study of confinement and the physics of flux tubes in this setup has a long history \cite{Schwinger:1962tp,Gross:1993hu,Dalley:1992yy,Gross:1995bp} and it received a renewed attention, see e.g. \cite{Cherman:2019hbq,Komargodski:2020mxz,Damia:2024kyt,Dempsey:2024alw}.}

However, there exist situations—most notably in the physically relevant case of QCD—in which the spectrum of bulk excitations is parametrically lighter than the strong--coupling scale. In such cases, confinement may still take place and the theory can support string--like excitations. A three--dimensional model exhibiting this hierarchy of scales, namely three--dimensional QED with Polyakov confinement, was recently analyzed in \cite{Aharony:2024ctf}. That study showed how the predictions of effective string theory are recovered below the mass gap, while also allowing for the computation of model--dependent corrections extending up to the strong--coupling scale. See also \cite{Caselle:2024zoh} for a numerical verification of such predictions.

One may also consider the qualitatively different situation in which the bulk theory is gapless. In this limiting case, the standard predictions of effective string theory are not expected to apply, since there exists no regime in which the dynamics can be effectively reduced to a purely two-dimensional description on the string worldsheet. The purpose of this work is to study an effective field theory (EFT) of massless "pions" where such phenomenon occurs and admits a perturbative approach. In particular we consider the four-dimensional $\mathbb{C}\mathbb{P} ^1$ non--linear sigma--model (NLSM) governing the dynamics of light degrees of freedom associated with the symmetry breaking pattern
\begin{equation}
SU(2) \,\, \to \,\, U(1)\,.
\end{equation}
Such IR effective theory is ubiquitous in high energy physics and describe, for instance, the low energy dynamics of the two-flavor Abelian-Higgs model and, remarkably, of Adjoint QCD with a single Dirac fermion \cite{Cordova:2018acb,DHoker:2024vii}.  

The theory supports solitonic string configurations protected by a topological charge classified by
$\pi_2(\mathbb{C}\mathbb{P}^1)=\mathbb{Z}$.
These objects can be viewed as the natural four-dimensional uplift of the two-dimensional Belavin-Polyakov instantons~\cite{Polyakov:1975yp}.
 From a modern perspective, the topological charge is associated to a global $U(1)$ 1-form symmetry \cite{Gaiotto:2014kfa}, hence protecting these excitations from decaying in the vacuum.

We will present a detailed analysis of these field configurations and the leading effects of their interactions with the four-dimensional pions. A natural and expected outcome of our analysis is that the behavior of observables associated to the string, such as the finite volume corrections to the vacuum energy and its characteristic width, differ considerably from the predictions of EST, due to the gapless bulk in which the string is embedded. Logically, the stark contrast with the $M_{gap}>0$ case stems from the emergent scale invariance of the leading order action of the NLSM. Specifically, we find that the spectrum of the theory in the presence of such string configurations continues to exhibit a continuum of states with arbitrarily low momenta, corresponding to bulk pion modes perturbed by the string background. In addition to these excitations, the spectrum contains extra massless Nambu-Goldstone bosons, associated with (a subset of) the classical moduli of the string solution and corresponding to broken symmetries. Remarkably, some of these moduli do not correspond to dynamical degrees of freedom in the infinite--volume limit and instead define genuine quantum parameters of the effective string theory.\footnote{One may also consider small symmetry--breaking perturbations that open a gap in the bulk spectrum. In this case, one can anticipate the emergence of light string bound states associated with the moduli related to the broken symmetries. We plan to investigate this scenario in a forthcoming work.
}

The rest of the paper is organized as follows. In Section~\ref{sec: 2} we review the classical properties of charge-$n$ string configurations in the $\bC\bP^1$ nonlinear sigma model, emphasizing the features that will be relevant for the subsequent analysis. In Section~\ref{sec: quadratic fluctuations} we introduce small quantum fluctuations around the classical background and determine the spectrum of string excitations. In particular, we identify which of the classical moduli give rise to dynamical degrees of freedom and which instead remain frozen in the infinite-volume limit. We then focus on the excited states and provide a detailed analytical and numerical analysis of their scattering with the soliton string. In Section~\ref{sec: 4} we compute the one-loop correction to the ground state energy of the charge-one closed string. We disentangle the contributions arising from finite size effects from those that survive in the infinite length limit, the latter being naturally interpreted as a quantum correction to the string tension. In Section~\ref{sec: 5} we determine the width of the string by computing the vev of the electric field perpendicular to the string worldsheet. In Section~\ref{sec: 6} we discuss possible ultraviolet completions of the $\bC\bP^1$ NLSM, focusing in particular on the $N_f=2$ Abelian-Higgs model and $SU(N_c)$ Adjoint QCD, and briefly comment on how the properties of the confining strings are expected to evolve along the RG flow. We conclude with three appendices containing technical details of the analytical and numerical computations presented in the main text.

\section{The $\mathbb{CP}^1$ model and its string-like solitons}\label{sec: 2}

Consider the four-dimensional nonlinear sigma model with target space $\bC\bP^1 \cong S^2$ defined by the action 
\begin{equation}
    \label{eq: CP1 action}
S= \frac{1}{2g^2} \int d^4 x \, \partial_\mu\vec n \cdot \partial^\mu\vec n 
\end{equation}
where $\vec n = (n_1,n_2,n_3)$ are dimensionless massless scalars subject to the constraint $|\vec n|^2 =1$  and $g$ is a dimensionful coupling constant.
A convenient parametrization of the target space, which automatically enforces the unit-norm constraint $|\vec n|^2 = 1$, is provided by the complex stereographic coordinates
\be
\vec n = \frac{1}{1+|\omega|^2}\left(\omega + \overline{\omega},\; -i(\omega - \overline{\omega}),\; |\omega|^2 - 1 \right),
\qquad
\omega = \frac{n_1 + i n_2}{1 - n_3},
\ee
where the north pole $\vec n = (0,0,1)$ is mapped to infinity in the $(\omega,\overline\omega)$ plane. In terms of these fields, the action becomes
\be\label{eq: CP1 action 2}
S = \frac{2}{g^2}\int d^4x\, \frac{\partial_\mu \omega\, \partial^\mu \overline{\omega}}{(1 + |\omega|^2)^2}\,.
\ee
 As a matter of fact, the theory  \eqref{eq: CP1 action} possess a conserved 2-form current
\be
\star J^{(2)} = n^*(\Omega)=\frac{1}{4\pi}\vec n \cdot \partial_\mu \vec n \times \partial_\nu \vec n \, dx^\mu \wedge dx^\nu \equiv \frac{i}{2\pi}\frac{d\omega \wedge d\overline{\omega}}{(1+|\omega|^2)^2}\,,
\ee
where $\Omega$ is the Kahler form of the target space and $n^*$ its pullback to spacetime through the pion field configuration. The above current is conserved as a consequence of the fact that the Kahler form is closed. Integrating $\star J^{(2)}$ over a closed two-dimensional surface measures the topological charge $n\in \bZ$ of a given string soliton configuration.

The manifest $O(3)$ global symmetry of \eqref{eq: CP1 action} is realized on the stereographic coordinates as follows. The $SO(3)$ subgroup acts by Möbius transformations,
\be\label{eq: global SU(2)}
    \omega \;\longmapsto\; \frac{a\,\omega + b}{-\overline{b}\,\omega + \overline{a}}\,, 
    \qquad
    \begin{pmatrix}
        a & b \\[2pt]
        -\overline{b} & \overline{a}
    \end{pmatrix} \in SU(2)
    \;\Rightarrow\; |a|^2 + |b|^2 = 1\,,
\ee
and only a $U(1)\subset SO(3)$, with $b=0$, is realized linearly. The remaining $\mathbb{Z}_2$ antipodal reflection $\vec n \mapsto -\vec n$ is implemented as
\be\label{eq: global reflection}
    \omega \;\longmapsto\; -\frac{1}{\overline{\omega}}\,.
\ee
The action \eqref{eq: CP1 action 2} gives rise to the following equations of motion
\begin{equation}\label{eq:eoms}
    \begin{split}
        &\partial_{\mu}\partial^{\mu} \omega - \frac{2 \overline{\omega}}{1+|\omega|^2}\partial_{\mu}\omega\partial^{\mu}\omega=0\, , \\ & 
       \partial_{\mu}\partial^{\mu} \overline{\omega} - \frac{2\omega}{1+|\omega|^2}\partial_\mu\overline{\omega}\partial^{\mu}\overline{\omega}=0\, .
    \end{split}
\end{equation}

The theory is known to support classically stable string-like solitons, namely solutions to the above equations that are independent of two of the four spacetime coordinates, carry non-trivial one-form symmetry charge and have finite tension.\footnote{Many of the statements reviewed below can be found in standard textbooks on the subject (see e.g.~\cite{Manton:2004tk}). Here, we focus on the properties of the classical string solutions that will be relevant for the quantum theory.
} Thus the 1-form symmetry is preserved and the theory is in a confined phase. 

These solutions can be naturally regarded as the four dimensional uplift of instanton excitations in the two dimensional rotor model \cite{Polyakov:1975yp}. For concreteness, we consider strings extended along the $x^3$-direction in space, therefore solutions to  \eqref{eq:eoms} of the form $\omega_{cl}=\omega_{cl}(x^1,x^2)$. The on-shell action evaluated on these solutions is therefore naturally interpreted as an energy density per unit length, {\it i.e.} a {\it tension}
\be
S_{cl}={\rm Vol}(\Sigma_w) \frac{2}{g^2}\int dx^1 dx^2 \frac{\partial_i \omega_{cl}\partial^i \omega_{cl}}{\left(1+|\omega_{cl}|^2\right)^2}={\rm Vol}(\Sigma_w) T_{cl}
\ee   
where $\Sigma_w$ denotes the worldsheet, namely the two dimensional manifold spanned by the $x^0$, $x^3$ coordinates. Finite tension solutions carry a non-trivial 1-form symmetry charge measured by integrating the 1-form symmetry current along the plane $\Sigma_\perp$ transverse to the string worldsheet. Moreover, their tension satisfies the following inequality \cite{Polyakov:1975yp}
\be\label{eq: BP bound}
T_{cl}\geq \frac{4\pi}{g^2}|Q(\Sigma_\perp)|\,.
\ee   
Henceforth, we will focus on configurations saturating the bound \eqref{eq: BP bound}. 

These are most easily described adopting complex coordinates $z=x^1+ix^2$, $\overline{z}=x^1-ix^2$ on $\Sigma_\perp$, in terms of which they reduce to holomorphic rational functions
\be\label{eq: rational map}
\omega_{cl}(z)=\frac{P(z)}{Q(z)} \,,
\ee
where $P(z)$ and $Q(z)$ are polynomials in $z$ with no common factor. Regarding $\omega_{cl}(z)$ as a map from $\Sigma_\perp$ (which by inclusion of infinity becomes topologically an $S^2$) to the target $S^2$, the topological charge $n$ is given by the degree of such a map, which is in turn given by  $|n|={\rm max} ({\rm deg}(P),{\rm deg}(Q))$,\footnote{
Given a map $\omega_{cl}: \, S^2 \to S^2$ of degreee $n$, it can be thought of as a $n$-covering of the target $S^2$. In practice, this means that a point $c$ on the target $S^2$ has $n$ preimages in the spacetime $S^2$. Accordingly, $|n|$ is determined by counting solutions to the algebraic equation
\begin{equation}
    \omega_{cl}(z)= c\, \rightarrow P(z)- c Q(z)= 0\, , 
\end{equation}
which are indeed $|n|=\text{max}(\text{deg}(P),\text{deg}(Q))$.} whereas the sign is determined by the relative orientations on the base and target spheres. Without loss of generality, we will restrict to $n\geq 0$.

In order to write down explicit solutions of this form, one needs to specify boundary conditions. Since the bulk is in a symmetry breaking phase, this amounts to choose a particular vacuum. A simple choice is
\be\label{eq: bc}
\vec n \xrightarrow{|z|\to \infty} \left(0,0,-1\right) \quad , \quad \omega \xrightarrow{|z|\to \infty} 0
\ee
while any other equivalent polarization is obtained via the action of the $O(3)$ global symmetry transformations \eqref{eq: global SU(2)}, \eqref{eq: global reflection}.

Given the boundary condition \eqref{eq: bc}, which immediately imply $\text{deg}(P)<\text{deg}(Q)$, the most general rational map of the form \eqref{eq: rational map} can be brought to the form
\begin{equation}\label{eq: general solution}
    \omega_{n}(z)= \frac{p_{1}z^{n-1}+ p_{2}z^{n-2}+...+ p_{n}}{z^{n}+q_{1}z^{n-1}+ p_{2}z^{n-2}+...+ q_{n}}\, ,
\end{equation}
for certain complex parameters $\{p_i,q_i\}$. The absence of common roots between $P(z)$ and $Q(z)$ leads to a single inequality satisfied by these coefficients. Therefore, the classical moduli space of solutions is identified with a certain $2n$-dimensional complex manifold $\cM_{2n}\subset \bC^{2n}$. Points along this manifold are naturally acted by the linearly realized $U(1)$ symmetry, whereas the broken generators map solutions satisfying different boundary conditions and do not define an action within $\cM_{2n}$. In addition, broken translations, rotations and dilatations also act within $\cM_{2n}$. As we will discuss in section \ref{sec: quadratic fluctuations}, some directions are lifted at the quantum level. 

As anticipated, the string tensions of these configurations saturate the bound \eqref{eq: BP bound}, which immediately implies the additive relation
\begin{equation}\label{eq: BPS relation}
    T_{n_1+n_2} = T_{n_1} + T_{n_2}\,.
\end{equation}
Consequently, there is no classical force between two separated strings, which accounts for the existence of the moduli described above.

For $n=1$ (i.e.~the fundamental unit-charge string) the moduli space is $\cM_{4} \cong \bC \times \bC^*$, and the corresponding classical configuration reads
\begin{equation}
    \omega_{n=1}(z) = \frac{\lambda\, e^{i\phi}}{z - z_0}\,,
\end{equation}
with $\lambda \in \bR_+$, $\phi \in [0,2\pi)$, and $z_0 \in \bC$. The parameter $z_0$ corresponds to the position of the string and is associated with the spontaneous breaking of Poincaré symmetry,
\begin{equation}
ISO(3,1) \;\rightarrow\; ISO(1,1) \times O(2)\,.
\end{equation}
The parameter $\lambda$ controls the classical width of the string while the angle $\phi$ sets the phase of the string, corresponding to the breaking of the $U(1)$ global symmetry of the bulk theory.

This configuration is rotationally symmetric around $z_0$, and without loss of generality we set $z_0 = 0$ by fixing the origin of the transverse coordinate plane. In contrast, strings with charge $n>1$ are generically not rotationally invariant, since a generic point in their moduli space corresponds to a configuration of multiple separated strings with lower charges.\footnote{Indeed the $4n$ real moduli can be understood as
the $2n$ positions, $n$ phases and $n$ sizes of the $n$ separated strings.} One can, however, consider the rotationally invariant charge-$n$ string configuration, described by the classical profile
\begin{equation}\label{eq: charge n rot sym}
    \omega^{rot}_{n}(z) = \left(\frac{\lambda\, e^{i\phi}}{z - z_0} \right)^n \,.
\end{equation}
It is clear that this configuration is restricted to a submanifold of the full moduli space.

\section{Quadratic fluctuations and the string spectrum}\label{sec: quadratic fluctuations}
The interactions of the string with the bulk massless degrees of freedom can be described, at least when fluctuations are small, by expanding the sigma model action around a given string solution. The first non-trivial contribution arises at quadratic order in fluctuations. This procedure results in a differential operator $P$ whose spectrum contains information about the scattering of the bulk pions off the string and, possibly, also bound states. Setting $\omega=\omega_0+g \eta$ with $\omega_0$ a generic field configuration, we get, for the quadratic part of the action\footnote{The prefactor $g$ is chosen so that the kinetic term for the fluctuation $\eta$ is canonically normalized.
}
\begin{equation}\label{eq: quadratic action}
\begin{split}
    S^{(2)}&= 2\int d^4 x\frac{1}{(1+|\omega_0|^2)^2}\Bigg[\left(\partial_\mu\eta - 2\frac{\overline{\omega}_0\partial_\mu\omega_0}{1+|\omega_0|^2}\eta\right)\left(\partial^\mu\overline{\eta} - 2\frac{\omega_0\partial^\mu\overline{\omega}_0}{1+|\omega_0|^2}\overline{\eta}\right)\\& + \frac{1}{(1+|\omega_0|^2)^2}\left(\eta^2\partial_{\mu}\overline{\omega}_0\partial^{\mu}\overline{\omega}_0+\overline{\eta}^2\partial_{\mu}\omega_0\partial^{\mu}\omega_0- 2 |\eta|^2\partial_{\mu}\omega_0 \partial^{\mu}\overline{\omega}_0\right) \Bigg] \, .
\end{split}
\end{equation}
To bring this to a physically transparent form we rescale $\eta=(1+|\omega_0|^2)\psi$ so that 
\begin{equation}\label{eq: S2 inner prod}
    S^{(2)}=\int d^4 x\begin{pmatrix}
        \overline{\psi}& \psi
    \end{pmatrix} P \begin{pmatrix}
        \psi\\ \overline{\psi}
    \end{pmatrix}\equiv \langle \Psi, P \Psi\rangle \, , \qquad \Psi=\begin{pmatrix}
        \psi\\ \overline{\psi}
    \end{pmatrix}
\end{equation}
where we introduced an inner product $\langle ,\rangle $ for complex doublets. The differential operator is 
\begin{equation}
    P= - D_\mu D^{\mu} \unit- X\, , \qquad D_{\mu}= \partial_{\mu}+A_{\mu} 
\end{equation}
where $A_{\mu}$ is the composite $U(1)$ gauge field
\begin{equation}
    A_{\mu}=  \frac{\omega_0\partial_\mu\overline{\omega}_0-\overline{\omega}_0\partial_\mu\omega_0}{1+|\omega_0|^2}
\end{equation}
and $X$ is the $2\times 2$ mass matrix
\begin{equation}\label{eq: X}
    X= \frac{2}{(1+|\omega_0|^2)^2}\begin{pmatrix}
\partial_\mu\omega_0\partial^\mu\overline{\omega}_0 & \partial_\mu\omega_0\partial^\mu\omega_0\\ \partial_\mu\overline{\omega}_0\partial^\mu\overline{\omega}_0& \partial_\mu\omega_0\partial^\mu\overline{\omega}_0
    \end{pmatrix}\, .
\end{equation}
Notice that the field strength of $A_{\mu}$ 
\begin{equation}\label{eq: F}
    F_{\mu \nu}= \frac{2}{(1+|\omega_0|^2)^2}\left(\partial_\mu\omega_0 \partial_\nu\overline{\omega}_0-\partial_\nu\omega_0 \partial_\mu\overline{\omega}_0\right)
\end{equation}
is proportional to the $1$-form symmetry current of the model and its conservation follows from the Bianchi identity. 

The fluctuation operator $P$ is manifestly self-adjoint with respect to the inner product defined in \eqref{eq: S2 inner prod}. In general, we can think of it as describing the motion of a particle in a magnetic field, subject to a potential $X$. 
We will be interested in expanding around the solitonic string configurations $\omega_0=\omega_n$, where it takes a diagonal form
\begin{equation}
    P_n= -\left( D_\mu D^{\mu} + 2\frac{\partial_\mu\omega_n\partial^{\mu}\overline{\omega}_n}{(1+|\omega_n|^2)^2}\right)\unit \, . 
\end{equation}


Remarkably, the operator $P_n$ is positive semi-definite, ensuring the absence of tachyonic instabilities around a given string soliton. In order to prove this, it is convenient to use complex coordinates $(z,\bar z)$ to parametrize the plane transverse to the worldsheet, thus having 
\be
F_{z\overline z}=[D_z, D_{\overline{z}}]=2\frac{\partial_z \omega_n \partial_{\overline z}\overline{\omega}_n}{\left(1+|\omega_n|^2\right)^2}
\ee
and therefore
\begin{equation}
    P_n=- \nabla^2_{\Sigma_w} -2\left(D_z D_{\overline{z}}+D_{\overline{z}}D_z -F_{z\overline{z}}\right)=- \nabla^2_{\Sigma_w} -4D_z D_{\overline{z}}\, .
\end{equation}
where $\nabla^2_{\Sigma_w}$ denotes the Laplacian on the worldsheet.
Since $D_z^{\dagger}= -D_{\overline{z}}$ we conclude that $P_n$ is positive semi-definite. As mentioned, this implies the absence of negative energy modes that would destabilize the string. Consequently, the lowest end of the spectrum is accounted for by the zero modes associated to the moduli discussed in the previous section. On top of these, and due to the string being embedded in a gapless bulk, there is a continuum of scattering states with no gap. 

In particular the zero modes are the solutions of
\begin{equation}
D_{\overline{z}}\psi(z,\overline{z})=0\, , 
\end{equation}
which take the form 
\begin{equation}
    \psi(z,\overline{z})= \frac{f(z)}{1+|\omega_n(z)|^2}\, , 
\end{equation}
where $f(z)$ (equivalently $\eta$ in \eqref{eq: quadratic action}) is an arbitrary holomorphic function.
As expected, wavefunctions obtained by taking derivatives of the general solutions \eqref{eq: general solution} with respect to any of the moduli are of this form.

\subsection{Moduli Quantization}

As a first approximation to study the dynamics of (extended) solitons, it is customary to assume that their motion happens on much larger scales than that of the fundamental fields of the theory, such that, at each fixed time, the solution looks the same (see for instance \cite{Rajaraman:1982is,Manton:2004tk} for a textbook approach). In practice, this amounts to endowing the moduli with a weak dependence on the worldsheet coordinates. In this sense, the moduli are promoted to two-dimensional fields, whose effective theory is naturally organized in a derivative expansion.   
In general, for the charge-$n$ string we write
\be
\omega_n=\omega_n(z, m_i(\sigma))\,,
\ee
where $m_i$ ($i=1,\ldots, 2n$) stands for the moduli and $\sigma_a$  denote the worldsheet coordinates, such that
\begin{equation}
\partial_{a}\omega_n=\sum_{i=1}^{2n}\frac{\partial \omega_n}{\partial m_i} \partial_{a}m_i \, .
\end{equation}
Expanding to quadratic order around the solution we have
\begin{equation}\label{eq: ZM kinetic terms}
\begin{split}
      S &= S_{cl} + \sum_{i,j}\int d^2\sigma \partial_{a}m_i \partial^{a}\overline{m}_j \left[\int d^2x\frac{\frac{\partial \omega_n}{\partial m_i} \frac{\partial \overline{\omega}_n}{\partial \overline{m}_j}}{(1+|\omega|^2)^2}\right] \,,
\end{split}
\end{equation}
where, in the expression within brackets, the $m_i$ are taken as constant. By worldsheet translation symmetry, this implies that we are expanding around arbitrary vacuum expectation values for these fields.

Given that, by definition, the action does not depend on any constant shift of the moduli, namely $S[\omega_n(m_i+\delta m_i)]=S[\omega_n(m_i)]$, and specializing the quadratic expansion \eqref{eq: S2 inner prod} to
\be
\Psi_{m_i}=\frac{1}{1+|\omega_n|^2} \begin{pmatrix}
        \frac{\partial\omega_n}{\partial m_i}\\ \frac{\partial\overline\omega_n}{\partial \overline{m}_i}
    \end{pmatrix} \, ,
\ee
one immediately verifies that
\be
P \Psi_{m_i} = 0 \, ,  
\ee
consistently with the fact that $\Psi_{m_i}$ is the wave function of a zero mode around the classical solution. 
Correspondingly the integrals determining the coefficients of the kinetic terms
\begin{equation}\label{eq:kinco}
    \int d^2x\frac{\frac{\partial \omega}{\partial m_i} \frac{\partial \overline{\omega}}{\partial \overline{m}_j}}{(1+|\omega|^2)^2}= \frac{1}{2}\langle \Psi_{m_i},  \Psi_{m_j}\rangle \, , 
\end{equation}
are precisely the norms of the zero mode wave functions. In particular, only normalizable zero modes are promoted to dynamical fields on the worldsheet. On the contrary, for a non-normalizable zero mode, the coefficient of the kinetic term in \eqref{eq: ZM kinetic terms} diverges and the corresponding modulus has {\it infinite inertia} \cite{Manton:2004tk}. Dynamically, any shift in their value costs an infinite amount of energy. Relatedly moduli associated to non-normalizable zero modes are frozen in the infinite volume limit, each value corresponding to a different superselection sector, specified by the boundary condition at infinity \cite{Intriligator:2013lca}.

\subsection{Spectral problem for rotationally symmetric strings}

We are interested in finding the eigenfunctions of the operator governing the quadratic spectrum, namely 
\begin{equation}
    P_n\psi= E \,\psi \, . 
\end{equation}
We can get rid of the worldsheet dependence by a plane wave decomposition $\psi(\sigma, z, \overline{z})= e^{i q\cdot \sigma}\psi(z,\overline{z})$, leading to the following two-dimensional spectral problem
\begin{equation}\label{eq: transverse diff eq}
   -4 D_z D_{\overline{z}}\psi(z,\overline{z})= k^2 \psi(z,\overline{z})\, .
\end{equation}
where $k^2\equiv E-q^2$.

For a generic solution, the above problem is quite hard to solve. However, it gets significantly simplified by restricting to solutions of the form \eqref{eq: charge n rot sym} enjoying axial symmetry.
In this case, it is convenient to use polar coordinates and perform a partial wave decomposition
\begin{equation}
    \psi(r,\theta)= \sum_{M\in \bZ} e^{i M \theta}\frac{\xi_M(r)}{\sqrt{r}}\, , 
\end{equation}
Defining the dimensionless coordinate $u=r/\lambda$ and energy $\kappa^2= \lambda^2 k^2$, the differential equation \eqref{eq: transverse diff eq} in the basis of partial waves takes the form
\be\label{eq: transverse spectral}
H \xi_M(u)=\kappa^2 \xi_M(u) 
\ee
with the Hamiltonian
\be
H=- \frac{d^2}{du^2} + \frac{1}{u^2}\left[M^2-\frac{1}{4}+ 4n\frac{M+n + u^{2n}(M-n)}{\left(u^{2n}+1\right)^2}\right]
\ee
In this formulation, the positivity of the fluctuation operator becomes manifest through the following factorization property
\be\label{eq: H factorization}
H= A^\dagger A
\ee
where
\begin{equation}
    A=\frac{d}{du}+ W(u) \, ,\qquad A^{\dagger}=-\frac{d}{du}+ W(u) \, , \qquad W(u) = -\frac{1}{u}\left(M+\frac{1}{2}\right) - \frac{2n}{u(u^{2n}+1)}  . 
\end{equation}
implying that for all partial waves the spectrum is non-negative. 

The factorization \eqref{eq: H factorization} immediately enables a reformulation of the spectral problem in terms of a {\it dual} Hamiltonian\footnote{\label{foot: dual spectra} Restricted to non-vanishing eigenvalues, {\it i.e.} $\kappa\neq0$, there is a bijective map between elements in the spectra of both Hamiltonians. More precisely,  if $\widetilde\Psi$ is an eigenfunction of $\widetilde H$, then $\Psi= A^{\dagger}\widetilde\Psi$ is an eigenfunction of $H$ with the same eigenvalue.}
\be\label{eq: dual H}
\widetilde H=A A^\dagger =-\frac{d^2}{du^2} + \frac{1}{u^2}\left[(M+1)^2-\frac{1}{4}+ 4n\frac{1+n+M}{u^{2n}+1}\right]\, , 
\ee
which will prove useful for some developments presented further below in this section. In passing, it is interesting to note that, for $M= -n-1$, the dual Hamiltonian becomes free, with eigenfunctions (imposing regularity at the origin) $\xi(u)= \sqrt{u}J_{|n|}(\kappa u)$. Moreover, it is straightforward to map this into an eigenfunction of $H$ and show that it also reduces to a free propagating wave, as it should.

Before tackling the resolution of the full spectral problem for axially symmetric strings, let us focus on the zero mode sector. It can be easily verified that the following functions
\begin{equation}
    \xi_M^{(0)}(u)= \frac{u^{M+2n+1/2}}{u^{2n}+1}\, . 
\end{equation}
satisfy $A \xi^{(0)}_M=0$, hence belonging to the kernel of the Hamiltonian $H$. In order to find the normalizable zero modes, we define the wave functions
\be
\Psi^{(0)}_M=\left(e^{i M\theta} \frac{u^{2n+M}}{u^{2n}+1}, e^{-i M\theta} \frac{u^{2n+M}}{u^{2n}+1}     \right)^T
\ee
and evaluate
\be
\left\langle \Psi^{(0)}_M, \Psi^{(0)}_{M'} \right\rangle= \delta_{M,M'} 
\frac{2\pi^2\lambda^2}{n^2}\frac{n+M+1}{\sin\left(\frac{\pi(-M-1)}{n}\right)}
\ee
We therefore find that normalizable zero modes exist only for
\be
-2n \leq M \leq -2
\ee
and we recall that we are working with $n>0$. \footnote{On the other hand, the co-kernel of $H$, namely the kernel of $\widetilde H$, is spanned by functions with partial waves of the form
$$
\widetilde\xi^{(0)}_M(u)= \left( u^{2n}+1\right) u^{-(2n+M+1/2)} 
$$
which, for $n>0$, have infinite norm for any $M$. On the contrary, if $n<0$ one finds normalizable zero modes of the dual Hamiltonian $\widetilde H$ for $0\leq M \leq 2|n|-1$, whereas no normalizable solutions exist for $H$. This is consistent with the fact that $H$ and $\widetilde H$ are exchanged by charge conjugation, which acts as $n\to -n$ on the string.  }

We conclude that, upon expanding around the charge-$n$ multistring configuration, there are $2n-1$ complex (equivalently $4n-2$ real) normalizable zero modes. Note that this result is valid for general charge $n$ solutions, even if obtained by working in the restricted subspace of axially symmetric ones, where the expectation value of most of these fields are set to zero.  

It is instructive to interpret this result in terms of the moduli for the charge-$n$ solution \eqref{eq: general solution}. At the classical level, the complex moduli space is $2n$-dimensional. According to the above analysis, $2n-1$ directions give rise to normalizable zero modes, hence leading to $2n-1$ complex scalars on the worldsheet, whereas the remaining one stays as a proper parameter, labeling superselection sectors in the effective theory. In order to address which of the original directions becomes a parameter, it is sufficient to consider a slightly more general solution parametrized as
\be 
\omega_n =\frac{\bar\lambda z^{n-1}+\lambda^{2n}}{z^n}
\ee 
which reduces to the axially symmetric one for $\bar\lambda\to 0$. In terms of this solution, it is straighforward to check that 
\be
\frac{1}{\left(1+|\omega|^2\right)^2}\frac{\partial\omega_n}{\partial\bar\lambda} \Big|_{\bar\lambda\to 0}\sim \Psi^{(0)}_{-1}
\ee
We conclude that the non-normalizable zero mode at $M=-1$ is associated to shifts of the $z^{n-1}$ power in the numerator of the general charge-$n$ multistring soliton. Notice that this mode behaves as an overall rescaling at large distances from the string, that is for large $|z|$, in agreement with a similar analysis presented in \cite{Intriligator:2013lca}.  For $n=1$, $\bar\lambda$ reduces to the parameter $\lambda$ which indeed parametrizes overall rescalings of the unit charge string, whereas, for $n>1$, the parameter $\lambda$ in \eqref{eq: charge n rot sym} leads to a normalizable zero mode.   

~

Now we move to the scattering states, namely solutions to the transverse spectral problem \eqref{eq: transverse spectral} with $\kappa^2>0$. As already emphasized, one may obtain the spectrum in terms of either the Hamiltonian $H$ or its dual $\widetilde H$. Even if, in both cases, an exact solution is out of reach, the equations take a simple form in the small and large $u$ regions
\begin{equation}
\begin{split}
      & u\gg 1 \, , \qquad -\xi''(u) - \frac{1}{u^2}\left(\frac{1}{4}-l^2\right)\xi(u)= \kappa ^2\xi(v)\, , \\
      & u\ll 1 \, , \qquad -\xi''(u) - \frac{1}{u^2}\left(\frac{1}{4}-(l+2n)^2\right)\xi(u)= \kappa ^2\xi(u)\, ,
\end{split}
\end{equation}
where $l=M$ for $H$ and $l=M+1$ for $\widetilde H$, and we momentarily omit the subscript in $\xi$ to avoid cluttering notation. 

In the large $u$ region the solution takes the form
\begin{equation}\label{eq:largeusol}
\begin{split}
      \xi(u) \simeq A_{n,l}(\kappa)\sqrt{u}\left(J_{|l|}(\kappa u)+ \tan(\delta_{l, n}(\kappa))Y_{|l|}(\kappa u)\right) \,,
\end{split}
\end{equation}
where we introduced the phase shift $\delta_{l, n}(\kappa)$. This solution describes a free propagating wave for a bulk pion after scattering off the string. The effects of the interactions among the pions and the string are encoded in the phase shift $\delta_{l,n}(\kappa)$, as well as the amplitude $A_{n,l}(\kappa)$. 

In turn, for small $u$ the regular solution behaves as 
\begin{equation}
    \xi(u) = \sqrt{u}J_{|l+2n|}(\kappa u)\sim u^{\frac{1}{2}+|l+ 2n|} \,,
\end{equation}
where we discarded the solution singular at $u=0$. Imposing regularity at the origin as an initial condition uniquely defines the phase shift $\delta_{l,n}(\kappa)$ but leaves the amplitude arbitrary. Using the relation between eigenfunctions of $H$ and $\widetilde H$ (see footnote \ref{foot: dual spectra}), it is straightforward to verify that the phase shifts derived from the two Hamiltonians coincide. 

In the next subsection we will implement a numerical computation, together with some analytical approximations, in order to determine the phase shifts $\delta_{l,n}(\kappa)$. As we will briefly review in Section \ref{sec: 4}, the intrinsic relation between the phase shift and the density of scattering states makes it a crucial ingredient for the computation of relevant observables such as the ground state energy.

\subsection{Phase shifts}
To efficiently compute the phase shifts numerically it is convenient to use the variable phase equation formalism \cite{calogero1967variable}. This amount to choose a parametrization of the wave function tailored for our purposes, namely
\begin{equation}
    \xi(u)= A_{n,l}(u,\kappa) \left[\cos(\delta_{n,l}(u, \kappa))\sqrt{u}J_{l}(\kappa u) + \sin(\delta_{n,l}(u,\kappa))\sqrt{u}Y_{l}(\kappa u)\right] \,,
\end{equation}
where $l=M$ or $l=M+1$ respectively for $H$ or $\widetilde H$.
Note that this is a parametrization of the exact solution, and the phase shift defined in \eqref{eq:largeusol} is recovered as 
\be\label{eq: phase shift form delta}
\delta_{n,l}(\kappa)=\lim_{u\to \infty}\delta_{n,l}(u,\kappa) \,.
\ee

Since the two Hamiltonians have isomorphic spectra we can choose either one, and we find it more convenient to solve for $\widetilde H$. So far we have two undetermined functions $A_{n,l}(u,\kappa)$ and $\delta_{n,l}(u,\kappa)$ but only one equation and one boundary condition $\xi(0)=0$. We are therefore free to impose a constraint, and we find it convenient to set
\begin{equation}
    \xi'(u) = A_{n,l}(u,\kappa)\kappa (\cos(\delta_{n,l}(u,\kappa))J'_{\nu}(\kappa u) + \sin(\delta_{n,l}(u,\kappa))Y'_{\nu}(\kappa u))\,,
\end{equation}
which allows to solve for $A_{n,l}(u,\kappa)$ in terms of $\delta_{n,l}(u,\kappa)$.\footnote{Solving \eqref{eq: xi eq} in the small $u$ region, one can check that there is a unique solution for $A_{n,l}$ and $\delta_{nl}$ that gives $\xi(0)=0$.} 

Making use of this ansatz, the eigenvalue equation 
\begin{equation}\label{eq: xi eq}
    -\xi''(u) + \frac{1}{u^2}\left(l^2-\frac{1}{4}\right)\xi(u) + 4n \frac{n+l}{u^2(u^{2n}+1)}\xi(u) = \kappa^2 \xi(u) \,,
\end{equation}
reduces to
\begin{equation}\label{eq: delta eq}
    \frac{d \delta_{n,l}(u, \kappa)}{du} = \frac{2 \pi n (n+l)}{u(u^{2n}+1)}\left[\cos(\delta_{n,l}(u, \kappa))J_{|l|}(\kappa u) + \sin(\delta_{n,l}(u, \kappa))Y_{|l|}(\kappa u)\right]^{2}\, .
\end{equation}
which is a much simpler equation than the full second order one. We need to solve it for $\delta_{n,l}(u,\kappa)$, supplemented by the boundary condition $\delta_{n,l}(0,\kappa)=0$, in order to extract the phase shift as in \eqref{eq: phase shift form delta}.

It will prove useful to attain a high energy approximation for the phase shift $\delta_{n,l}(\kappa)$, also known as Born approximation. In fact, assuming that $\tan(\delta_{n,l}(u,\kappa))$ vanishes as $\kappa\to\infty$, the equation \eqref{eq: delta eq} gets simplified to 
\be
 \frac{d \delta_{n,l}(u, \kappa)}{du}= \frac{2 n \pi(n+l)}{u(u^{2n}+1)}J_{|l|}(\kappa u)^{2} \, ,
\ee
and yields
\be
\delta_{n,l}(\kappa)=\delta_{n,l}(\infty)+ 2 n \pi(n+l)\int_{0}^{\infty}\frac{J_{|l|}(\kappa u)^{2}}{u(u^{2n}+1)}\, .
\ee
The above solution is only valid for large values of $\kappa$, hence it will be instructive to derive an expansion in powers of $\kappa^{-1}$. 
In particular, for $n=1$, the integral converges for all values of $l\neq 0$ and can be done analytically, obtaining
\begin{equation}\label{eq: Born}
\begin{split}
    \delta_{1,l}(\kappa) &= \delta_{1,l}(\infty)+2 \pi (l+1)\left(\frac{1}{2|l|}- I_{|l|}(\kappa)K_{|l|}(\kappa)\right)-\pi\left(1+\frac{1}{|l|}\right) \\
    &\simeq\delta_{1,l}(\infty) -\frac{\pi(l+1)}{\kappa} + O(1/\kappa^3)\, .
\end{split}
\end{equation}

The total phase shift is obtained by summing over all partial waves $l\in\bZ$. As per the discussion below \eqref{eq: dual H}, we recall that, for $l=-n$, the Hamiltonian \eqref{eq: dual H} becomes free and, consequently, the phase shift satisfies $\delta_{n,-n}=0$. In view of this, it turns out to be convenient to sum symmetrically around this point. We therefore define the partial sums 
\begin{equation}
    \Delta_{n,l}(\kappa)=\delta_{n,l}(\kappa) + \delta_{n, -l-2n}(\kappa)\, .
\end{equation}
One advantage of the partial sums defined above is that, form the Born approximation, and for $n=1$, the slower decay $\sim \kappa^{-1}$ in the phase shifts exactly cancels
\begin{equation}\label{eq: Delta Born}
    \Delta_{l,1}(\kappa)=\delta_{1,l}(\kappa) + \delta_{1, -l-2}(\kappa)\simeq \delta_{1,l}(\infty)+\delta_{1,-l-2}(\infty)+ O(1/\kappa^3)\, ,
\end{equation}
significantly improving the convergence of the numerical calculations presented below.  

From now on we will focus on the case $n=1$. Besides rendering a simpler problem, this object enables the exploration of the full space of parameters. Indeed, as per our discussion about the moduli quantization, the overall complex modulus $\lambda$ remains as a genuine parameter of the effective theory at the quantum level. This enables studying how the relevant observables depend on this parameter,\footnote{Due to the preserved $U(1)$ symmetry, the phase of $\lambda$ drops and generic observables will only depend on its magnitude. } as we will explore in sections \ref{sec: 4} and \ref{sec: 5}. On the other hand, for axially symmetry solutions with $n>1$, $\lambda$ is not a tunable parameter but represents a vacuum expectation value of a dynamical field which has to be stabilized, for instance, by minimizing the ground state energy. Moreover, since rotationally symmetric solutions only represent a one-dimensional submanifold of the full moduli space associated to $n>1$ solutions, it is not even guaranteed for global minima to exist within this restricted manifold. 

 \begin{figure}
     \centering
     \includegraphics[scale=0.32]{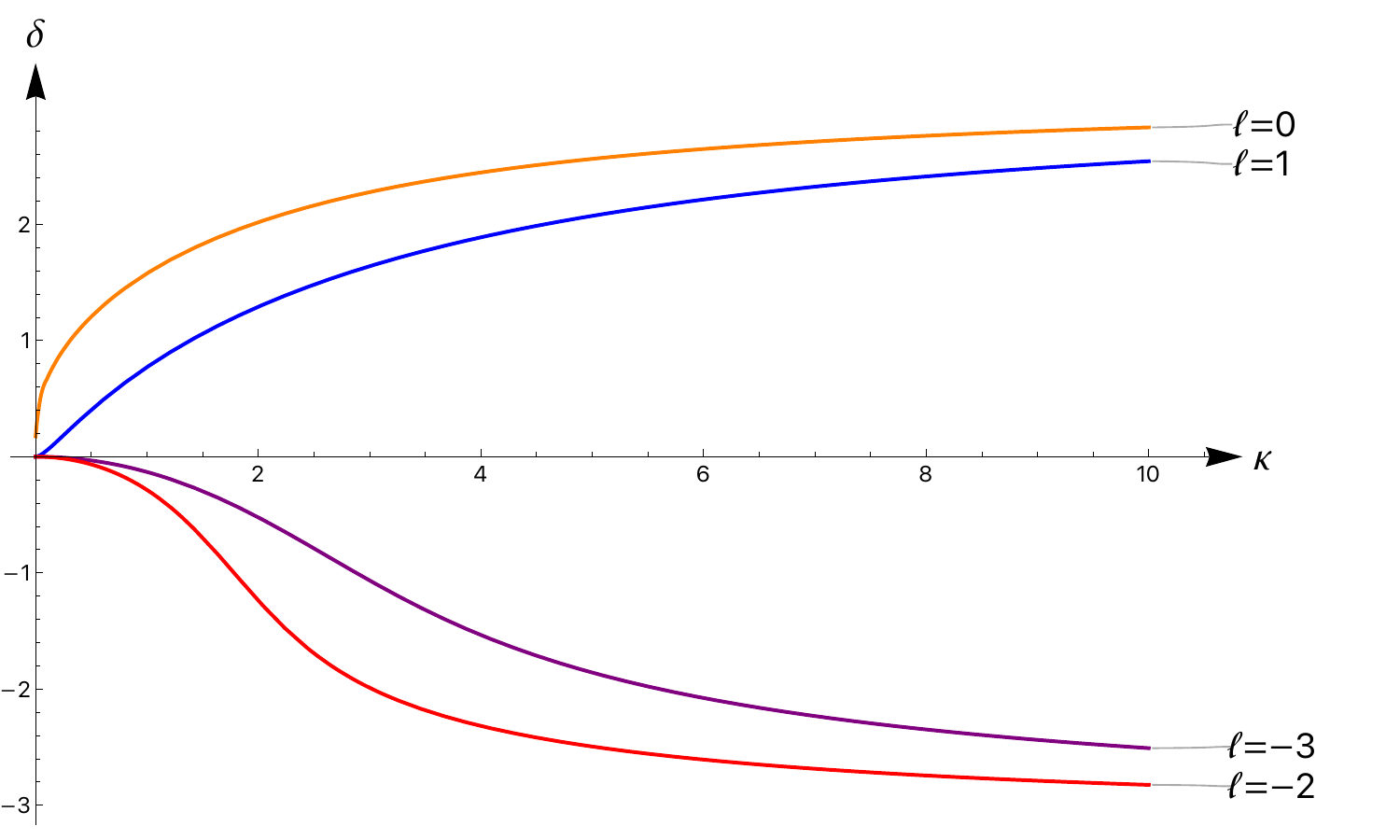}
      \includegraphics[scale=0.32]{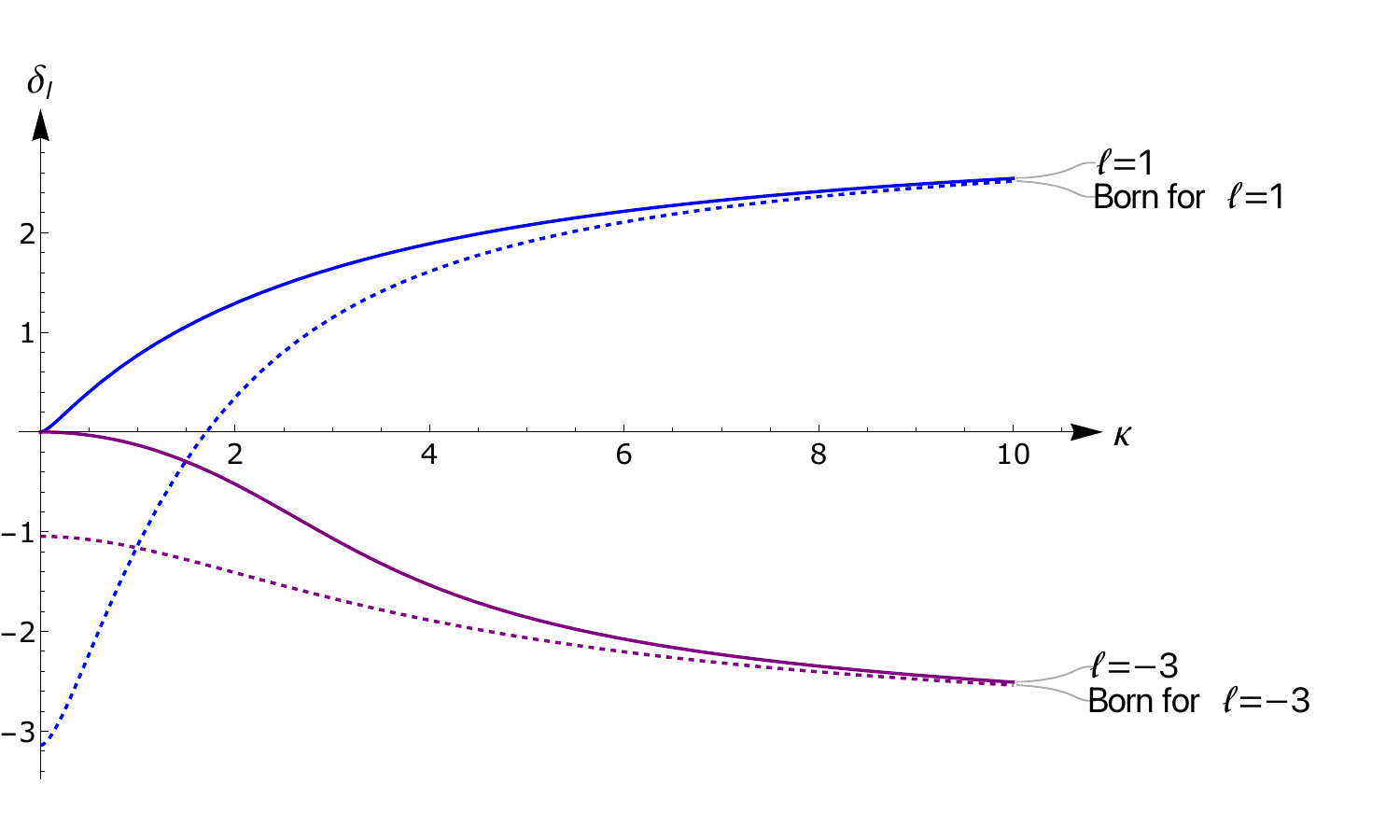}
     \caption{Left: phase shifts for $l=-3,-2,0,1$. Right: Comparison with the Born approximation, which is accurate at large $\kappa$.}
     \label{fig:PS_1}
 \end{figure}

With this in mind, we solve the phase equation \eqref{eq: delta eq} numerically and extract the phase shift for several values of $l$, always including its image with respect to reflections around $l=-1$ (which has vanishing phase shift), that is the partial wave with $-l-2$. We find the asymptotic values
\begin{equation}
    \delta_{1,l}(\infty)=\begin{cases}\pi\, \quad \text{for} \quad l\ge 0 \\-\pi\, \quad \text{for} \quad l\le -2 \end{cases} \, ,
\end{equation}
ensuring that the partial sums $\Delta_{1,l}(\kappa)$ vanish at infinity conforming with the Born approximation. We plot the first few phase shifts and compare them to the Born approximation in Fig.\ref{fig:PS_1}. 

\begin{figure}
    \centering
    \includegraphics[scale=0.4]{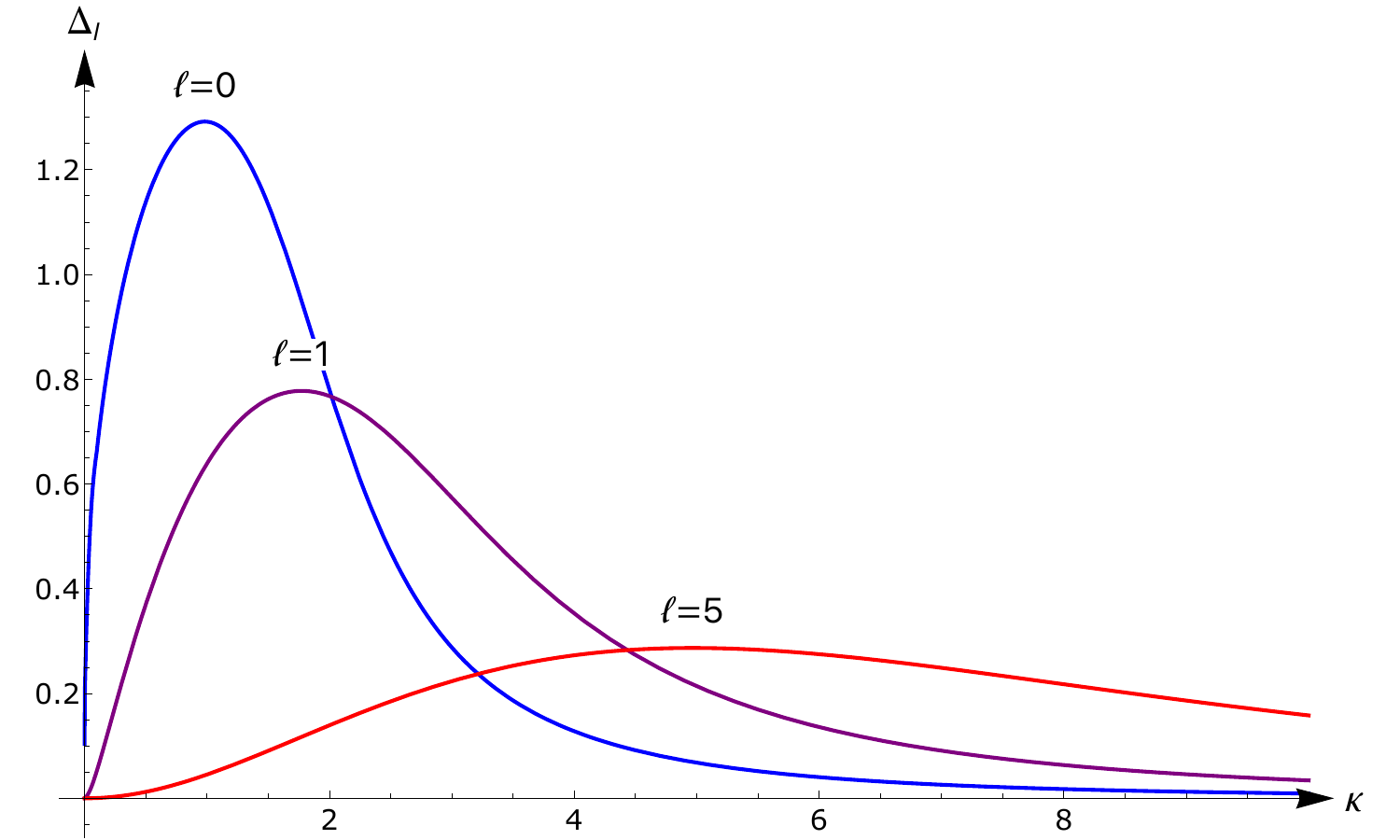}
    \caption{Partial sums $\Delta_{1,l}(\kappa)$ for $l=0,1,5$ .}
    \label{fig:PartialSums_1}
\end{figure}

We show the numerical results for the partial sums with $l=0, 5, 10$ in Fig.\ref{fig:PartialSums_1}. For generic values of $l$, we find that the partial sums $\Delta_{1,l}(\kappa)$ display the following general features
\begin{itemize}
    \item For $l\neq 0$, $\Delta_{1,l}(\kappa)$ vanishes as a positive power $\kappa^{\alpha(l)}$ for $\kappa\to 0$, with $\alpha(l)$ increasing as $l$ increases. On the other hand, as predicted by the high energy approximation, it decays as $1/\kappa^3$ at infinity. 
    
    \item The case of the $s$-wave, namely $l=0$, is special. As a matter of fact, the partial sum $\Delta_{1,0}(\kappa)$ vanishes as $1/\log(\kappa)$ at $\kappa=0$, hence significantly slower than the other partial waves (this is also the behavior evidenced for the phase shift itself). This implies that the effectively two-dimensional scattering cross section is divergent in the IR, as the density of states satisfies $\rho\sim d\delta/d\kappa$ (see Section \ref{sec: 4}). This is a well known feature pertaining to the scattering in two dimensions \cite{Friedrich:2015obs}.
    
    \item $\Delta_{1,l}(\kappa)$ attains its maximal value for an energy $\kappa_{\text{max}}(l)$ that grows linearly with $l$, $\kappa_{\text{max}}(l)\sim  l$, while the height of the maximum decreases as $\Delta_{1,l}(\kappa_{\text{max}}(l))\sim 1/l$. 
    
    \item Defining a width of $\Delta_{1,l}(\kappa)$ as the value $\kappa_w$ at which $\Delta_{1,l}(\kappa)$ attains, roughly, $1/10$ of its maximal value, we find that $\kappa_w$ also scales linearly with $l$. Therefore the phase shifts of higher angular momentum compensate their decrease in size by spreading out in energy. Quantitatively, one can verify that area below the curves does not depend on $l$. 
\end{itemize}

We proceed to the implementation of our numerical results for the partial sums $\Delta_{1,l}$ in order to compute the total phase shift. For a given value of $\kappa$, a direct implementation would require summing over partial waves up to some maximum value $l_{max}$
\be\label{eq: direct}
{\rm Direct:} \qquad \delta_1(\kappa;l_{max})=\sum_{l=0}^{l_{max}} \Delta_{1,l}(\kappa) \,,
\ee
subsequently asking for good convergence in $l_{max}$, namely
\be\label{eq: error}
{\rm E}(l_{max})=\frac{\left|\delta_1(\kappa;l_{max}+1)- \delta_1(\kappa;l_{max})\right|}{|\delta_1(\kappa;l_{max})|} \ll 1\,.
\ee
However, due to the properties of $\Delta_{1,l}(\kappa)$, for any choice of $l_{\max}$ the inequality \eqref{eq: error} is inevitably violated for $\kappa \gtrsim \sqrt{l_{\max}}$. In practice, we have computed $\Delta_{1,l}$ numerically up to $l_{\max}=100$, which ensures good convergence of the direct total phase shift \eqref{eq: direct} only up to $\kappa \sim 10$. This limitation is problematic, since the efficiency of the direct numerical implementation deteriorates rapidly as $\kappa$ increases.    

\begin{figure}
    \centering
    \includegraphics[scale=0.32]{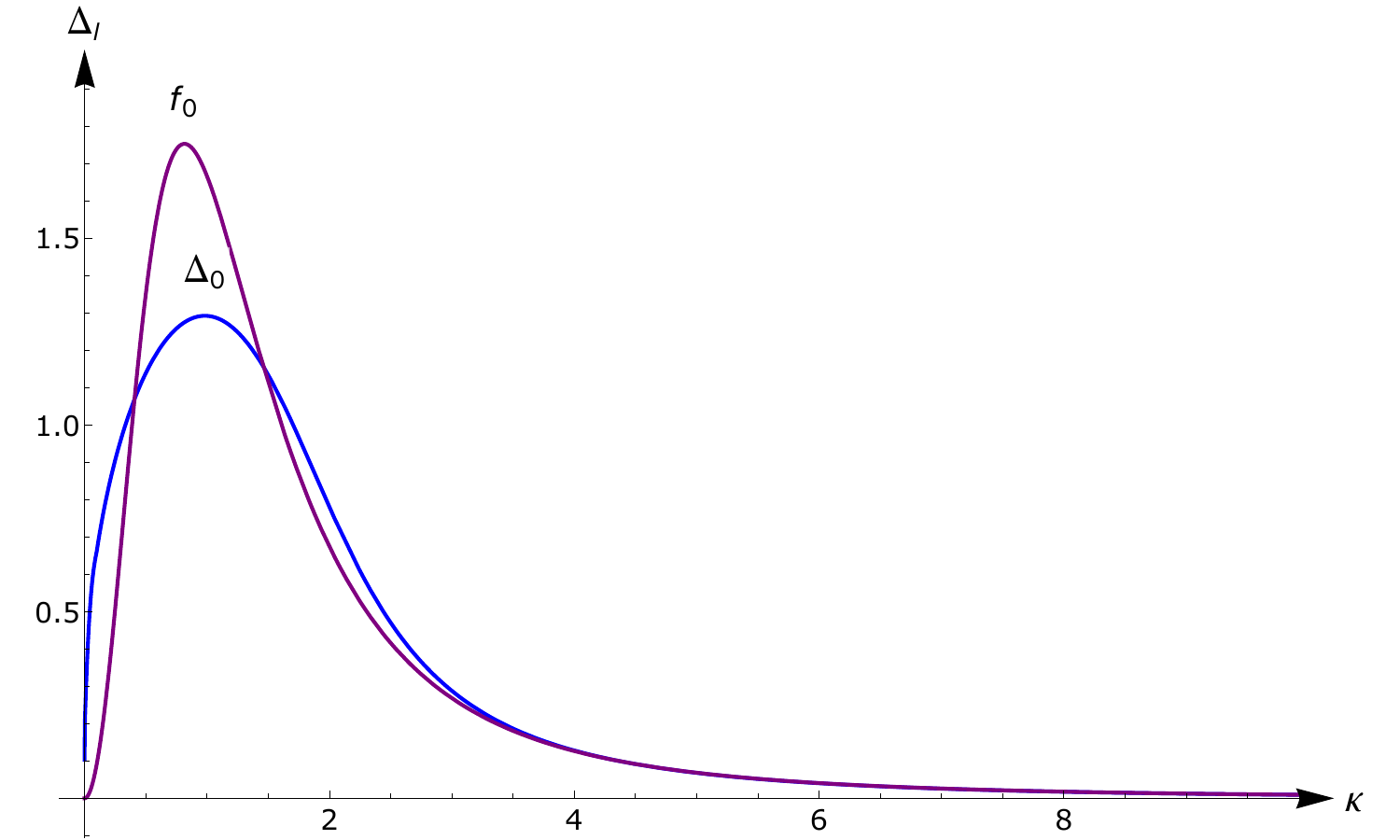}
    \includegraphics[scale=0.32]{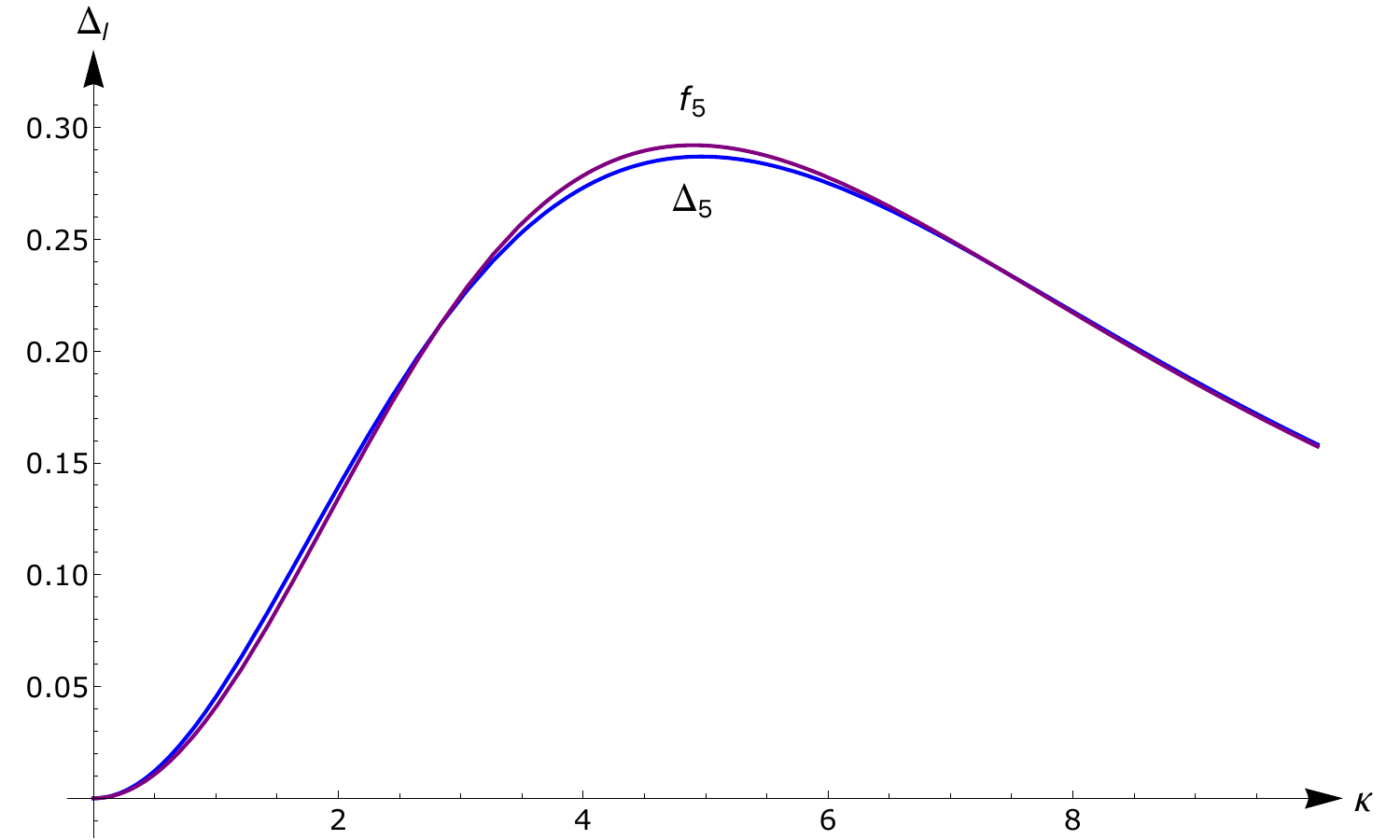}
     \includegraphics[scale=0.32]{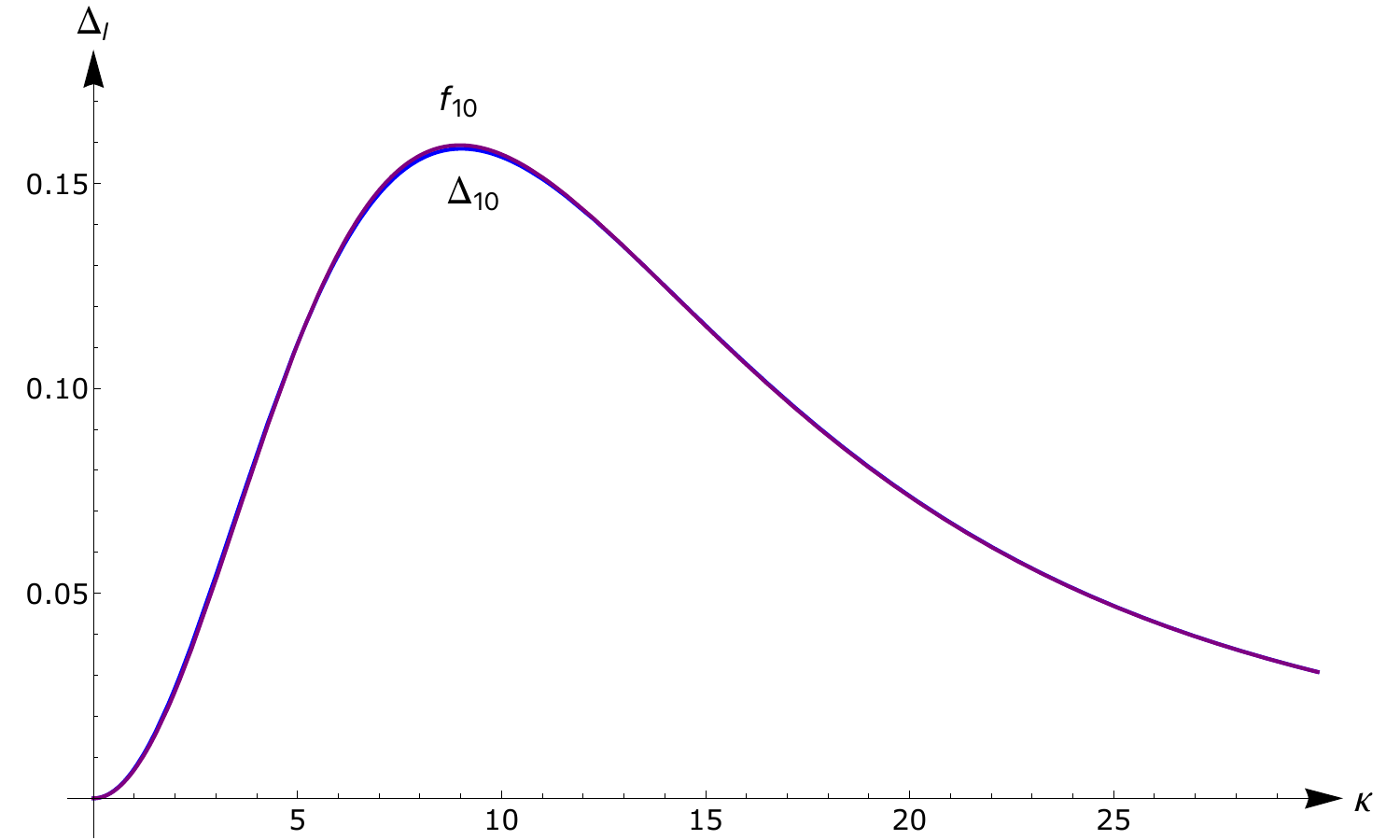}
    \caption{Comparing the fit models with the numerical results for $l=0, 5, 10$.}
    \label{fig:Fitl0}
\end{figure}

In order to overcome this issue, we propose an improved implementation of the numerical results, namely
\be\label{eq: improved}
{\rm Improved:} \qquad \delta_1(\kappa; l_{max})= \sum_{l=0}^{l_{max}} \Delta_{1,l}(\kappa) + \sum_{l_{max}+1}^{\infty}f_l(\kappa) \,,
\ee
where $f_l(\kappa)$ is a function that fits the numerical result for large values of $l$. In particular, we find the the rational function
\begin{equation}\label{eq: fitting function}
    f_{l}(\kappa) = 3\pi(l+1)^2 \frac{\kappa^2}{((l+1)^2+\kappa^2)^{5/2}} \,,
\end{equation}
provides and excellent fit of the numerical data already starting at $l\sim 10$. Moreover, it satisfies all the general properties observed for the exact partial sums $\Delta_{1,l}$. On the one hand, it decays as $\kappa^{-3}$ for $\kappa\to \infty$, reproducing the behavior predicted by the Born approximation. In addition, it can be easilly verified that its maximum satisfies ${\rm Max}[f_l]\sim l$ and its width, defined as before, decays as $l^{-1}$. Finally, we notice that the integral
\be
\int_0^{\infty} d\kappa f_l(\kappa) =\pi\, ,
\ee
is independent of $l$, consistently with our general observations. 
More quantitatively, we find that the difference between the fitting function and the numerical curve lies below $10^{-4}$ and $10^{-5}$ respectively at $l\sim 10$ and $l\sim 100$, for all values of $\kappa$. 

Therefore, we make use of this fitted model function \eqref{eq: fitting function} within our improved numerical determination of the total phase shift \eqref{eq: improved}, that is
\begin{equation}
    \delta_{1}(\kappa) = \sum_{l=0}^{l_{max}}\Delta_{1,l}(\kappa) + 3\pi\sum_{l=l_{max}+1}^{\infty}\frac{\kappa^2 (l+1)^2}{((l+1)^2+\kappa^2)^{5/2}}\, .
\end{equation}
In passing, we also notice that $f_l(\kappa)\sim l^{-3}$ for $l\gg \kappa$, rendering the above sum absolutely convergent. 
\begin{figure}
    \includegraphics[scale=0.32]{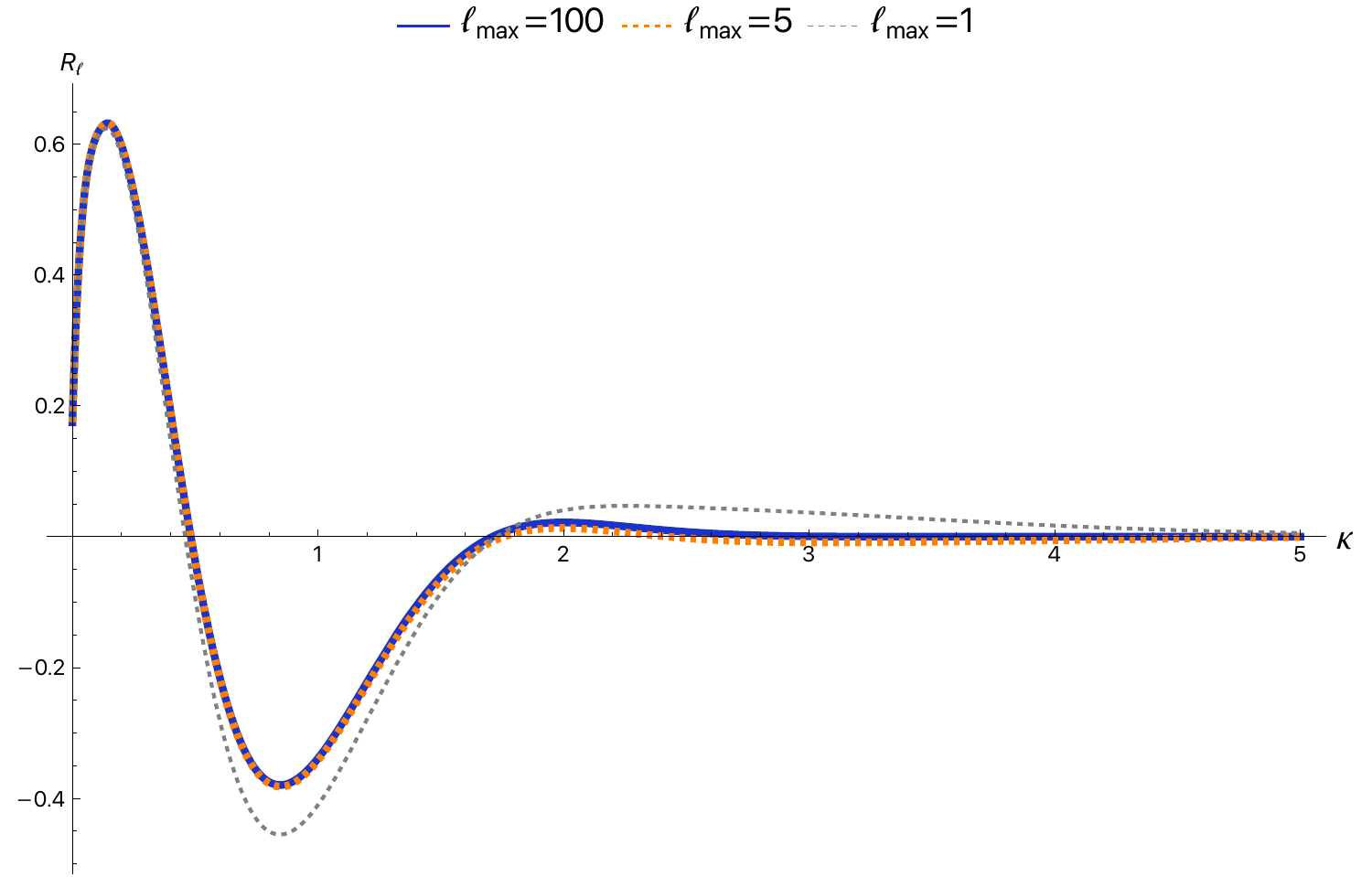}
     \includegraphics[scale=0.32]{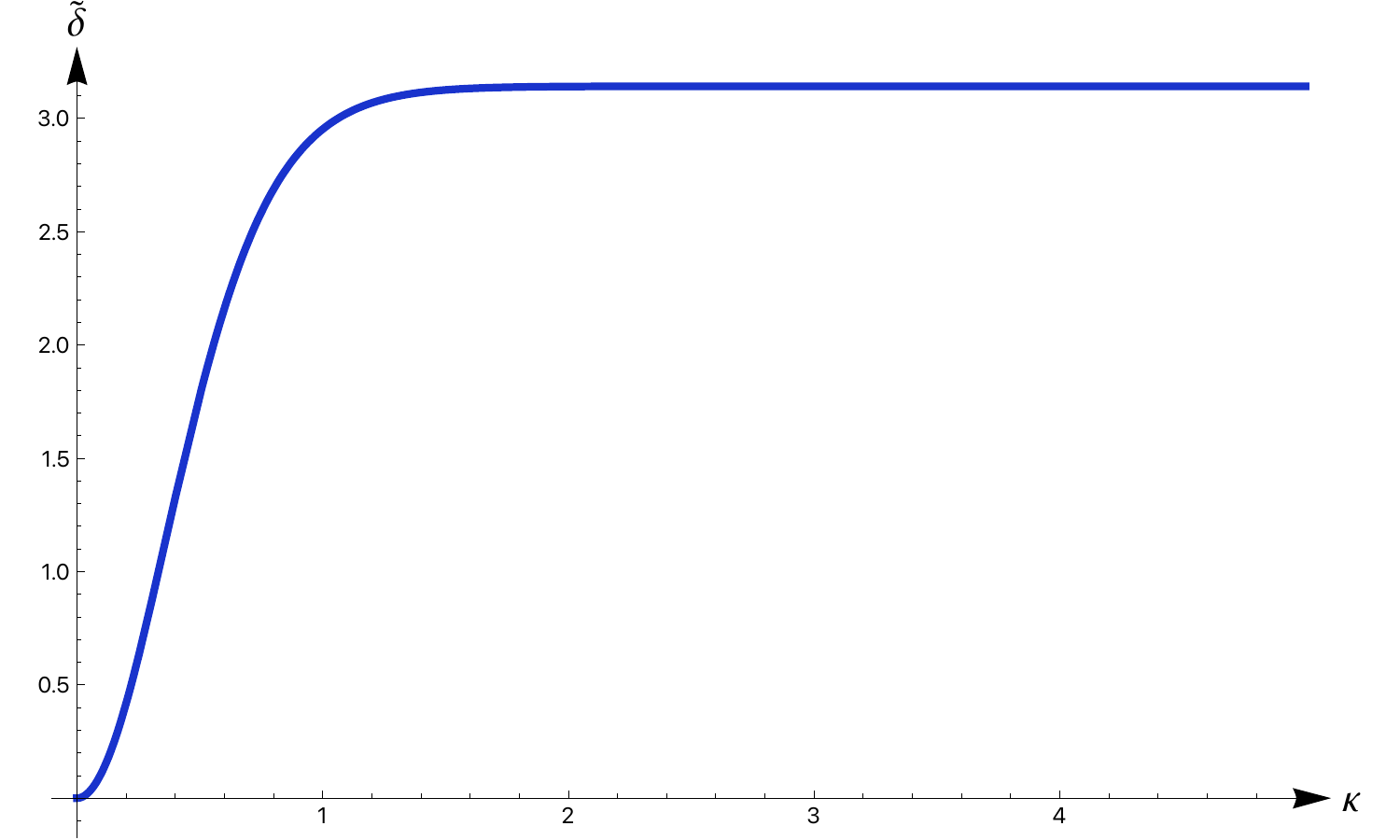}
   \caption{Left: the function $R_{l_{max}}(\kappa)$ for $l_{max}=1, 5, 100$. Right: the analytic function $\widetilde{\delta}_1(\kappa)$.}
    \label{fig:RLdeltat}
\end{figure}

It is instructive to compare our result with similar numerical computations already present in the literature. In particular, Belavin-Polyakov soliton solutions in the three-dimensional version of our model were studied in \cite{Moss:1999xs, PhysRevB.61.2819}. In order to compare with their findings we first notice that our result for $\delta_{1}(\kappa)$ does not vanish at infinity, but rather rapidly approaches $\pi$ for $\kappa\to \infty$. Subtracting this asymptotic value and computing the integral of the total phase shift we correctly reproduce the numerical value for the Belavin-Polyakov soliton Casimir energy reported in \cite{Moss:1999xs, PhysRevB.61.2819} (see Appendix \ref{app:PSints}).

As it is going to prove useful in later sections, we present here a convenient decomposition of the total phase shift 
\be
  \delta_1(\kappa) =  \sum_{l=0}^{l_{max}}(\Delta_{1,l}(\kappa)- f_{l}(\kappa)) + \sum_{l=0}^{\infty}f_{l}(\kappa)\equiv R_{l_{max}}(\kappa) + \widetilde{\delta}_1(\kappa)\, .
\ee
In this way, all the dependence on $l_{max}$ is encoded in the reminder function $R_{l_{max}}(\kappa)$. This quantity turns out to converge much faster than the sum over partial sums $\Delta_{1,l}$, making it more efficient for numerical evaluation. 

\begin{figure}
     \centering
      \includegraphics[scale=0.4]{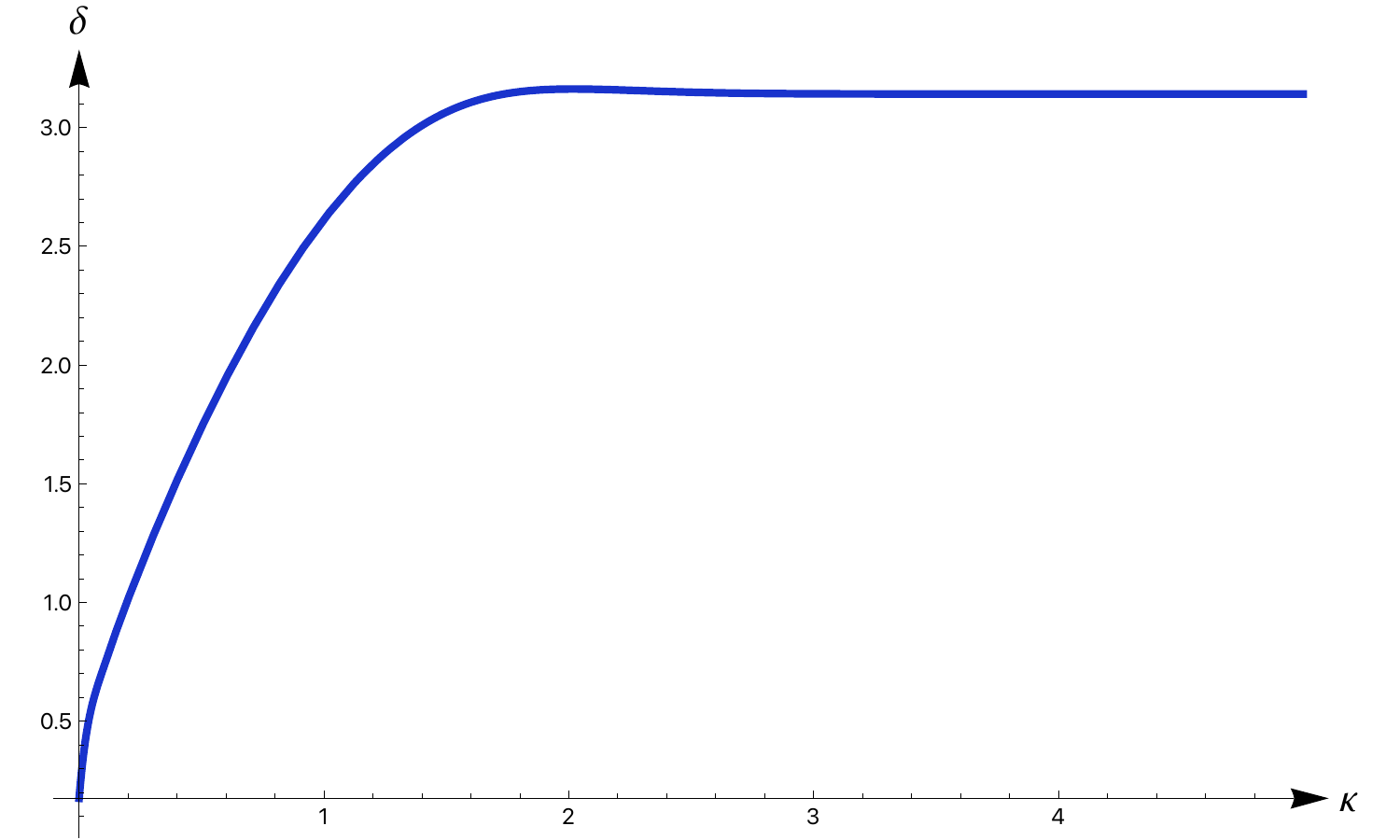}
   \caption{The total phase shift as a fuction of $\kappa$ and $l_{max}=100$. }
    \label{fig:deltatot}
\end{figure}

On the other hand, the analytical, $l_{max}$-independent, component admits the following representation
\begin{equation}
\begin{split}
      \widetilde{\delta}_1(\kappa)&= 3\pi \sum_{l=1}^{\infty}\frac{\kappa^2 l^2}{(l^2+\kappa^2)^{5/2}}= \frac{3\pi}{2}\sum_{m\in\bZ}\int_{-\infty}^{\infty} dl \frac{\kappa^2 l^2}{(l^2+\kappa^2)^{5/2}} e^{2i\pi m l}\\& = \pi + (2 \pi)^2 \frac{d}{d\kappa}\sum_{m=1}^{\infty} \kappa^2 m K_{1}(2 \pi \kappa m)= \pi + 4\pi^2\frac{d}{d\kappa}\int_0^{\infty}dt \cosh(t) \frac{\kappa^2 e^{2 \pi \kappa \cosh(t)}}{(e^{2 \pi \kappa \cosh(t)}-1)^2}\, , 
\end{split}
\end{equation}
which allows to easily extract its asymptotic behavior at high energies
\begin{equation}
     \widetilde{\delta}_1(\kappa)\simeq \pi + O(e^{- 2 \pi \kappa}\kappa^{3/2})\, .
\end{equation}
We plot the functions $R_{l_{max}}(\kappa)$ and $\widetilde{\delta}_1(\kappa)$ in Fig.\ref{fig:RLdeltat} and the total phase shift in Fig.\ref{fig:deltatot}. In particular we notice the fast decay of $R_{l_{max}}(\kappa)$ for $\kappa>1$, together with its rapid convergence in $l_{max}$.

\section{Ground state energy}\label{sec: 4}

The aim of this section is to make use of the previous analysis to compute the
one–loop correction to the ground–state energy of the fundamental string. This quantity is obtained by performing the path integral over the
quadratic fluctuations:
\begin{equation}
    e^{-L_t  E_{n}}
    = e^{-S(\omega_{\rm cl})}
    \int D\psi\, D\overline{\psi}\;
    \exp\!\left[
        -
        \int d^{4}x\,
        \begin{pmatrix}
            \overline{\psi} & \psi
        \end{pmatrix}
        P_{n}
        \begin{pmatrix}
            \psi \\[2pt] \overline{\psi}
        \end{pmatrix}
    \right],
\end{equation}
where $L_t$ is the length of the time direction and the classical contribution is
\begin{equation}
    S(\omega_{\rm cl})
    = \frac{4\pi |n|}{g^{2}}\,\text{Vol}(\Sigma_{w})\,.
\end{equation}
As customary, we consider closed strings with Vol$(\Sigma_w)=L_tL$. Since the operator $P_{n}$ is positive definite, the path integral is
well defined. The operator $P_n$ above encompasses the dynamics of bulk pions as well as their interactions with the charge-$n$ string soliton. In order to isolate the latter contribution from the pure four dimensional bulk physics, we subtract the one-loop ground–state energy in the absence of the string, denoted by $P_0$. The one–loop string vacuum energy thus yields
\begin{equation}
    E_{n}
    = \frac{4\pi |n|}{g^{2}}\,L
      + \frac{1}{L_t}\Tr\!\left( \log P_n - \log P_{0} \right) \equiv \frac{4\pi |n|}{g^{2}}\,L
      +\delta E_n.
\end{equation}
where the leading linear potential, typical for confined objects, arises from the classical action.
To make contact with the analysis of the previous section, we express the
operator trace in its spectral representation:
\begin{equation}\label{eq: 1loop E}
  \frac{1}{L}\delta E_n
    = (4n-2)\frac{1}{2}\int \frac{d^{2}q}{(2\pi)^{2}}
        \log\!\left(q^{2}\right)
      + \int \frac{d^{2}q}{(2\pi)^{2}}
        \frac{d^{2}k}{(2\pi)^{2}}\,
        \bigl[\rho_{n}(k) - \rho_{0}\bigr]\,
        \log\!\left(q^{2}+k^{2}\right),
\end{equation}
where $\rho_{n}$ denotes the density of states in the presence of the charge-$n$
string, and we have explicitly separated the contribution of the
zero modes localized on the string from that of the continuum bulk spectrum. In equation \eqref{eq: 1loop E}, $\vec q$ denotes the momenta along the
worldsheet.\footnote{ It is important to emphasize that the above expression is formal, in the sense that the integral over the spatial component of $\vec q$ is really a sum over the eigenvalues of the momentum operator on the prescribed worldsheet manifold, as we will show below for a circular string of length $L$.}
On the other hand, $\vec k$ is the momentum transverse to the string.  
For axially symmetric string configurations the density of states
depends only on the magnitude $k=|\vec k|$ of the transverse momentum, so that
$\rho_{n} = \rho_{n}(k)$. Finally, the factor $(4n-2)$ counts the number of dynamical Goldstone bosons
arising from the quantization of the moduli space associated with the charge $n$ string  (see the discussion in Section~\ref{sec: quadratic fluctuations}).

The quantity \eqref{eq: 1loop E} is the main subject of study of this section and we devote the rest of it to computing the leading contributions within our effective theory, valid for sufficiently long strings. On general grounds, such contributions will arise in two conceptually distinct types, depending on their behavior as the string length $L$ grows large. On the one hand, there is an extensive piece that asymptotes to a constant in the limit $L\to \infty$, hence encoding the one-loop corrections to the effective string tension. In addition, there are important finite size effects due to the fluctuations of the localized Goldstone fields, leading to universal contributions such as the Luscher term, as well as the more interesting non-universal contributions originated from the scattering of the bulk pions off the string. Due to the absence of a characteristic mass scale in the bulk effective field theory, we expect these effects to be controlled by the ratio $\lambda/L$, with $\lambda$ being proportional to the overall size of the string solution along the transverse directions. 

We will proceed to isolate these two types of contributions and compute them separately. Before delving into that, let us briefly review how the density of states is naturally connected to the phase-shift (see also \cite{Aharony:2024ctf}).  To this aim it is instructive to first consider the system in finite volume, eventually taking the infinite volume limit at the end. For rotationally symmetric strings, we can consider the radial problem in a finite box of size $R$, imposing the Dirichlet boundary condition $\psi_l(R)=0$ on each partial wave. In terms of the rescaled variables introduced in the previous section, the asymptotic form of the solution far away from the string reads  
\begin{equation}
\begin{split}
     \psi_l(u)&= A\sqrt{u}(J_{|l|}(\kappa u) + \tan(\delta_{n,l}(\kappa))Y_{|l|}(\kappa u))\\
     &=A\sqrt{ \frac{2}{\pi \kappa}} \frac{\cos\left(\kappa u -\frac{|l|\pi}{2}-\frac{\pi}{4}-\delta_{n,l}(\kappa)\right)}{\cos(\delta_{n,l}(\kappa))}
\end{split}
\end{equation}
and therefore the boundary condition imposes the following quantization condition
\begin{equation}
    \kappa_n R -\frac{|l|\pi}{2}-\frac{\pi}{4}-\delta_{n,l}(\kappa_n)= \left(m+\frac{1}{2}\right)\pi\, \, \, , \, \,\, m\in \bZ \,.
\end{equation}
As $R\to \infty$,  the eigenvalues bunch together and distribute according to the density of states  

\begin{equation}
     \frac{d m}{dk}= \frac{R}{\pi}- \frac{1}{\pi}\frac{d\delta_{n,l}(\kappa)}{d\kappa}\, ,
\end{equation}
where the first term $R/\pi$ is related to density of states in absence of the string. In turn, the total density of states results from summing over the contributions of each partial wave $l\in\bZ$. Finally, recalling the definition of the rescaled transverse momentum $\kappa= \lambda k$, the two dimensional density of states and the phase shift are related as
\begin{equation}\label{eq: density of state}
    \Delta m= \int \frac{d^2k}{(2\pi)^2} \rho_n(k) = \int dk \frac{k}{2\pi}\rho_n(k) \;\Longrightarrow\;\rho_n(k)-\rho_0 = -\lambda^2 \frac{2}{\kappa}\frac{d\delta(\kappa)}{d\kappa}=:\lambda^2\widehat{\rho}_n(\kappa) \, .
\end{equation}

\subsection{Infinite $L$ contribution and 1-loop string tension}

We begin by computing the extensive piece of the one-loop the vacuum energy density which modifies the string tension as
\begin{equation}
    T_n = \frac{4\pi}{g^2}|n| + \delta T_n\,.
\end{equation}

Within the typical framework of EST, where only the localized Goldstone fields are taken into account, all higher order interactions lead to finite size effects. Accordingly, the string tension plays the role of a physical cutoff and does not get corrected at any order. On the contrary, as we will verify below, shifts in the effective string tension originate from the interactions with the gapless bulk degrees of freedom.

This quantity is deeply intertwined with the renormalization of the bulk effective field theory. In fact, the integrals over the worldsheet momenta lead to UV divergent contributions to the one-loop string tension and require to be cancelled by appropriate counterterms.    
As usual in effective field theory, the structure of these divergences demands to include higher order terms in the effective action,\footnote{The renormalization of the pion decay constant $g^{-1}$ does not play any role at this order and can be reabsorbed by a simple redefinition of the classical tension.} in particular the following four-derivative interactions for the bulk pions\footnote{There is an additional coupling allowed by the symmetries of the problem. However, such coupling is proportional to the classical equations of motion, hence being redundant at one-loop.}
\be\label{eq: 4 derivative S}
S_{4\partial} =\int d^4 x \left[  y_1 \frac{(\partial_\mu \omega\partial^\mu\bar\omega)^2}{\left(1+|\omega|^2\right)^4}+ y_2  \frac{(\partial_\mu \omega\partial^\mu\omega)(\partial_\mu \bar\omega\partial^\mu\bar\omega)}{\left(1+|\omega|^2\right)^4}\right] \,.
\ee
It is striaghtforward to verify that the first term yields a non-vanishing contribution to the tension when evaluated on the charge-$n$ solution
\be\label{eq: tension counterterm}
\delta T_n^{4\partial} =  \frac{y_1}{\lambda^2}  B_n  \qquad , \qquad B_n\equiv  \lambda^2\int d^2 x_\perp \frac{(\partial_\mu \omega_n\partial^\mu\bar\omega_n)^2}{\left(1+|\omega_n|^2\right)^4} > 0 \,,
\ee
whit $\lambda$ the modulus determining the overall size of the string soliton.\footnote{For $n>1$, $B_n$ may depend on the additional moduli pertaining to the classical string configuration.}

Let us pause here to emphasize that four-derivative couplings \eqref{eq: 4 derivative S} belong to the bulk effective field theory and, as such, get renormalized by the bulk quantum effects, independently of the presence of the string soliton. In particular, this implies that the divergencies carried by these couplings are completely determined by the consistency of the four-dimensional theory and {\it cannot be further adjusted}. It is convenient to decompose
\be
y_1=y_1^r+\delta y_1 \,,
\ee
where the counterterm $\delta y_1$ encodes the singular behavior and $y_1^r$ is the finite renormalized coupling. Within dimensional regularization, the bulk interactions induce the following structure for the counterterm 
\be
\delta y_1=-\frac{\beta_{y_1}}{2}\left(\frac{1}{\epsilon}+\gamma_E-\log(4\pi)\right)\equiv -\frac{\beta_{y_1}}{2\bar\epsilon} \,,
\ee
where $d=4+2\epsilon$. To leading order in perturbation theory, $\beta_{y_1}$ is a positive constant that determines the running of the renormalized coupling
\be\label{eq: renormalized y1}
y_1^r = y_1(\mu_0)+\beta_{y_1} \log\left(\mu/\mu_0\right) \,,
\ee
with $\mu_0$ some reference UV scale that one may fix by other considerations. 

In Appendix \ref{app:betaf} we extract the coefficient $\beta_{y_1}$ from the expansion of the heat kernel, applied to the four-dimensional differential operator $P_n$, obtaining
\be\label{eq: beta y1}
\beta_{y_1}=\frac{1}{6\pi^2} \,\, .
\ee

Back into the computation of the string tension, a stringent consistency check of our effective theory arises from the fact that the one-loop divergencies entering in  string tension must be precisely cancelled by the counterterm contribution \eqref{eq: tension counterterm}, which is in turn fixed by \eqref{eq: beta y1}. By the explicit computation presented below, we will verify that such cancelation does take place and, relatedly, the one-loop corrected string tension is independent of the sliding scale $\mu$, as it should. 

A further important remark concerns the scheme dependence of this quantity. As a consequence of the procedure just described, we will find that it actually depends on our choice for the reference scale $\mu_0$ in \eqref{eq: renormalized y1}.
Therefore, the tension is not an intrinsic observable of the string. For soliton strings with topological charge $n>1$, a genuinely scheme-independent quantity is instead the combination
\begin{equation}
\Delta T_{n} = T_n - |n|T_1  \, ,
\end{equation}
which measures the stability of a charge-$n$ string from decaying into $n$ strings of unit charge. Classically this quantity vanishes as a consequence of \eqref{eq: BPS relation}. However, quantum corrections will generically spoil this relation and will determine the actual stability.

Keeping this in mind, in what follows we will focus on determining the tension of the unit charge string. By the above considerations, this computation stands as a valuable consistency check of our effective description.

From \eqref{eq: 1loop E} we find
\begin{equation}
\begin{split}
    \delta T_1&= \int \frac{d^2 q}{(2\pi)^2}\left[\log(q^2) + \int_0^{\infty} \frac{d \kappa}{2\pi}\widehat{\rho}_1(\kappa)\kappa \log\left(q^2 + \frac{\kappa^2}{\lambda^2}\right)\right]\\ & =\int \frac{d^2 q}{(2\pi)^2}\log(q^2)\left[1 + \int_0^{\infty} \frac{d \kappa}{2\pi}\widehat{\rho}_1(\kappa)\kappa \right] + \int \frac{d^2 q}{(2\pi)^2}\int_0^{\infty} \frac{d \kappa}{2\pi}\widehat{\rho}_1(\kappa)\kappa\log\left(1+ \frac{\kappa^2}{q^{2}\lambda^2}\right)\,.
\end{split}
\end{equation}
Using \eqref{eq: density of state} we see that the log-divergent term on the worldsheet is multiplied by
\begin{equation}
    1 + \int_0^{\infty} \frac{d \kappa}{2\pi}\widehat{\rho}_1(\kappa)\kappa  = 1 - \frac{1}{\pi}(\delta_1(\infty) - \delta_{1}(0))=0 \, ,
\end{equation}
where we used that $\delta_1(\infty)=\pi$.\footnote{This cancellation actually takes place also for charge-$n$ strings, where the presence of $(4n-2)$ NG bosons leads to a different asymptotic behaviour of the phase shift. Nonetheless, the divergences cancel in the same manner.} This is the manifestation of the fact that string localized NG bosons do not contribute to the string tension.

Therefore we are left with 
\begin{equation}
     \delta T_1=-\frac{1}{\pi}\int \frac{d^2 q}{(2\pi)^2}\int_0^{\infty} d \kappa\frac{d \delta_1(\kappa)}{d\kappa}\log\left(1+ \frac{\kappa^2}{q^{2}\lambda^2}\right)\, . 
\end{equation}
While the asymptotic behaviour of the phase shift ensures that the integral over $\kappa$ is finite, it is easy to verify that it diverges logarithmically in the worldsheet momentum $q$. We employ dimensional regularization, setting $d=2+2\epsilon$ and $1/\overline{\epsilon}=1/\epsilon+ \gamma_E-\log(4\pi)$ and obtain
\begin{equation}
\begin{split}
     \delta T_1&=-\frac{1}{\pi}\int_0^{\infty} d \kappa\frac{d \delta_1(\kappa)}{d\kappa}\int \frac{d^{d} q}{(2\pi)^d}\mu^{-2\epsilon}\log\left(1+ \frac{\kappa^2}{q^{2}\lambda^2}\right)\\ & = -\frac{1}{\pi \lambda^2}\int_0^{\infty} d \kappa\frac{d \delta_1(\kappa)}{d\kappa}\left[-\frac{\kappa^2}{4\pi\overline{\epsilon}}+\frac{\kappa^2}{4\pi}\log(\mu^2\lambda^2)+\frac{\kappa^2}{4\pi}(1-\log\kappa^2)\right]\, , 
\end{split}
\end{equation}
which can be rewritten more compactly as
\begin{equation}
    \delta T_1= \frac{\beta}{\lambda^2\overline{\epsilon}} - \frac{\beta}{\lambda^2}\log(\mu^2\lambda^2) +\frac{\alpha}{\lambda^2}\, , 
\end{equation}
where 
\begin{equation}
    \begin{split}
     &\alpha= \frac{1}{4\pi^2}\int_0^{\infty}d\kappa\frac{d\delta_1(\kappa)}{d\kappa}\kappa^{2} (\log(\kappa^2)-1)\\
     &\beta= \frac{1}{4\pi^2}\int_0^{\infty}d\kappa\frac{d\delta_1(\kappa)}{d\kappa}\kappa^{2} \, ,
\end{split}
\end{equation}
are real and finite numbers which can be computed numerically.

The renormalized string tension is attained by includying the contribution of the counterterm \eqref{eq: tension counterterm} and yields 
\begin{equation}
    T_1
    = \frac{1}{g^2}\left(
    4\pi
    + \frac{g^2 \alpha}{\lambda^2}
    + \frac{g^2}{\lambda^2}\left(
        \frac{\beta}{\overline{\epsilon}}
        - \beta \log(\mu^2 \lambda^2)
        - \frac{B_1}{12\pi^2}\frac{1}{\overline{\epsilon}}
        + B_1\, y_1^r(\mu)
    \right)
    \right)\, .
\end{equation}
As explained, a consistent renormalization requires
\begin{equation}\label{eq: value of beta}
    \beta = \frac{B_1}{12\pi^2} = \frac{1}{9\pi}\, ,
\end{equation}
hence imposing a nontrivial constraint on the phase shift describing the scattering of bulk pions off the string. Remarkably, even without an analytic expression for the phase shift, one can show on general grounds that this identity must hold (see Appendix \ref{app:betaf} for the proof). Moreover we can check the validity of \eqref{eq: value of beta} numerically (see Appendix \ref{app:PSints}).

Carrying through the divergence cancellation and using \eqref{eq: renormalized y1}, we find that the renormalized string tension depends on the renormalization scheme through the explicit dependence on the reference scale $\mu_0$. As it is standard in this type of theories (see for instance \cite{Aharony:2024ctf} for a similar analysis in three-dimensional QED), a sensible choice amounts to fix $\mu_0$ in terms of the characteristic scale of the string, namely $\mu_0=\lambda^{-1}$, in order to avoid large logarithms which might invalidate the perturbative expansion.

Implementing this choice leads to the following renormalized string tension
\begin{equation}
    T_1
    = \frac{1}{g^2}\left(4\pi+ \frac{g^2 \alpha}{\lambda^2}+ \frac{g^2}{\lambda^2} B_1y_1(\lambda^{-1})
    \right)\, ,
\end{equation}
Note that it depends on the 4-derivative couplings at the scale of the string, and the particular value of $\alpha$ is not physically meaningful, as it can be reabsorbed by $\mu_0\to \mu_0\exp{(\alpha/\beta)}.$ 

Finally notice that the actual expansion we get is in powers of $\left(\frac{g}{\lambda}\right)^2$, implying that the EFT expansion is valid for $\lambda \gg g$.

\subsection{Finite $L $ contribution}

Let us proceed to the evaluation of the finite–size effects. To this aim we define the quantity
\begin{equation}
   \delta T^{fs}_n(L ) = 
   \frac{1}{L }\,\delta E_n(L )
   -
   \lim_{L  \to \infty}\frac{1}{L }\,\delta E_n(L )\, .
\end{equation}
Note that $\delta T_n^{fs}(L )$ is scheme independent and does not depend on the
counterterms required to renormalize the string tension.

The zero-mode contribution is universal and it gives only finite-size contributions. These are obtained by summing over the discrete momentum along the string
\begin{equation}
     \left(\delta T^{fs}_{n}(L )\right)_{\text{z.m.}}= (2n-1)\frac{1}{L }\sum_{m\in \bZ}\int_{-\infty}^{\infty} \frac{dq}{(2\pi)}\log\left(q^{2}+\left(\frac{2\pi m}{L }\right)^2\right)\, .
\end{equation}
Since the intergral in UV divergent, we need to regularize it. As opposed to the one-loop string tension, the regularized quatity has a finite limit and does not require renormalization. We choose zeta-funtion regularization such that
\begin{equation}
    \left(\delta T^{fs}_{n}(L )\right)_{\text{z.m.}}=\partial_s\left[ (2n-1)\frac{1}{L }\sum_{m\in \bZ}\int_{-\infty}^{\infty} \frac{dq}{(2\pi)}\left(q^{2}+\left(\frac{2\pi m}{L }\right)^2\right)^s\,\right]_{s=0}\, .
\end{equation}
A straightforward computation then leads to the result
\begin{equation}
   \left(\delta T^{fs}_{n}(L )\right)_{\text{z.m.}}= -\frac{(2n-1)\pi }{3L ^2} \, ,
\end{equation}
which is the universal Luscher correction predicted by EST \cite{LUSCHER1981317}.

As expected, more interesting contributions originate from interactions with the bulk modes. Within the same regularization scheme we find
\begin{equation}
\begin{split}
     \left(\delta T^{fs}_{n}(L )\right)_{\text{bulk}}
     &=  \partial_s\Bigg[
     \int_{-\infty}^{\infty} \frac{dq}{2\pi} 
     \int_0^{\infty} \frac{d \kappa}{2\pi}\, \widehat{\rho}_n(\kappa)\, \kappa 
     \Bigg(
     \frac{1}{L } \sum_{m\in \bZ} 
     \left(q^2 + \left(\frac{2\pi m}{L }\right)^2 + \frac{\kappa^2}{\lambda^2}\right)^{s} \\
     &\qquad\qquad\qquad\qquad\qquad\qquad
     - \int_{-\infty}^{+\infty} \frac{dl}{2\pi} 
     \left(q^2 + l^2 + \frac{\kappa^2}{\lambda^2}\right)^{s} 
     \Bigg)
     \Bigg]_{s=0}\\
&= \partial_s \Bigg[\frac{1}{4\pi \Gamma(-s)} 
      \int_0^{\infty} \frac{d \kappa}{2\pi} \widehat{\rho}_n(\kappa) \, \kappa 
      \Bigg(
      \left(\frac{2\pi}{L }\right)^{2s+2}
      \sum_{r\in\bZ} \int_0^{\infty} dx\, x^{-s-2} 
      e^{- \frac{\pi^2 r^2}{x} - \left(\frac{\kappa L }{2\pi \lambda}\right)^2 x} \\
      &\qquad\qquad\qquad\qquad\qquad\qquad
      - \Gamma(-1-s) \left(\frac{\kappa}{\lambda}\right)^{2+2s}
      \Bigg)\Bigg]_{s=0}\, ,
\end{split}
\end{equation}
where, in going to the second line above, we performed the integrals over $q$ and $l$ and made use of the following identities
\begin{equation}
    \left(\left(\frac{2\pi m}{L }\right)^2 + \frac{\kappa^2}{\lambda^2}\right)^{s+1/2}
    = \left(\frac{2\pi}{L }\right)^{2s+1}
    \frac{1}{\Gamma\left(-s-\frac{1}{2}\right)}
    \int_{0}^{\infty} dx\, x^{-s-3/2}\,
    e^{- m^2 x \left(\frac{\kappa L }{2\pi \lambda}\right)^2 x}\, 
\end{equation}
and 
\begin{equation}
    \sum_{m\in\bZ}e^{-m^2 x}= \sqrt{\frac{\pi}{x}}\sum_{r\in\bZ}e^{- \frac{\pi^2 r^2}{x}}\,.
\end{equation}

In this form, one can check that the infinite $L $ term is precisely canceled by the $r=0$ term of the sum. In turn, the integral over $x$ can be expressed in terms of the Bessel function $K_{s+1}(r v)$ with $v=\kappa L/\lambda$. In terms of this variable and implementing the realtion between the density of states and the phase shift we arrive to the final expression for the leading finite size correction
\begin{equation}\label{eq: final fs bulk}
      \left(\delta T^{fs}_{n}(L)\right)_{\text{bulk}}= -\frac{2}{\pi^2L ^2}\int_0 ^{\infty}d v\,  v \, \delta_{n}\left(\frac{\lambda}{L }v\right)  \sum_{r=1}^{\infty} K_{0}\left(r v\right) \, .
\end{equation}

Some comments on this result are in order:

\begin{itemize}

\item Making use of the integral representation of the Bessel function we can replace the sum by an integral
\begin{equation}
    \sum_{r=1}^{\infty} K_0(r v)
    = \sum_{r=1}^{\infty} \int_0^{\infty} dt\, e^{- r v \cosh t}
    = \int_0^{\infty} dt\, \frac{1}{e^{v \cosh t} - 1}\,,
\end{equation}
where exchanging the sum and the integral is justified since the sum is absolutely convergent. 
Moreover, for $v \to 0$, we have
\begin{equation}
    \int_0^{\infty} dt\, \frac{1}{e^{v \cosh t} - 1} 
    \simeq \frac{1}{v} \int_0^{\infty} \frac{dt}{\cosh t} 
    = \frac{\pi}{2 v}\,.
\end{equation}
Hence, the integrand in  \eqref{eq: final fs bulk} is finite at $v = 0$ and exponentially suppressed for large $v$.

\item As already emphasized in Section~\ref{sec: quadratic fluctuations}, for the unit charge string the modulus 
$\lambda$ is frozen and stands as a genuine dimensionful parameter. 
Focusing on this case and collecting the contributions due to the NG bosons and the bulk pions, we can write our result as a "modified" Luscher term 
\be
    \delta T_{1}^{fs}\left(L/\lambda\right)= -\frac{\pi}{3L^2}\left[1+J\left(L/\lambda\right)\right] \,,
\ee
with
\be
J(x)=\frac{6}{\pi^3}\int_0^{\infty} dv \, v \, \delta_1(v/x) \int_0^{\infty} dt \frac{1}{e^{v \cosh t}-1} \,.
\ee
The integrals defining $J$ can be computed numerically as we report in Appendix \ref{app:PSints}. Notice that the EST result is recovered in the decoupling limit ($L/\lambda\to \infty$) since the function $J$ vanishes as
\be
J(x)\sim \frac{1}{\log\left(x\right)} \quad , \quad x\to \infty\,,
\ee
see Appendix \ref{app:PSints}.\footnote{The limit $L/\lambda \rightarrow \infty$ should be understood as considering very long strings, since the EFT we are considering is valid for $\lambda \gg g$.} This behavior can be understood from the fact that, in the deep infrared, the $\bC\bP^1$ NLSM reduces to a theory of free scalar fields, so that in this regime the string dynamics decouples and the energy receives contributions only from the NG modes. 
\begin{figure}
\centering
   \includegraphics[scale=0.4]{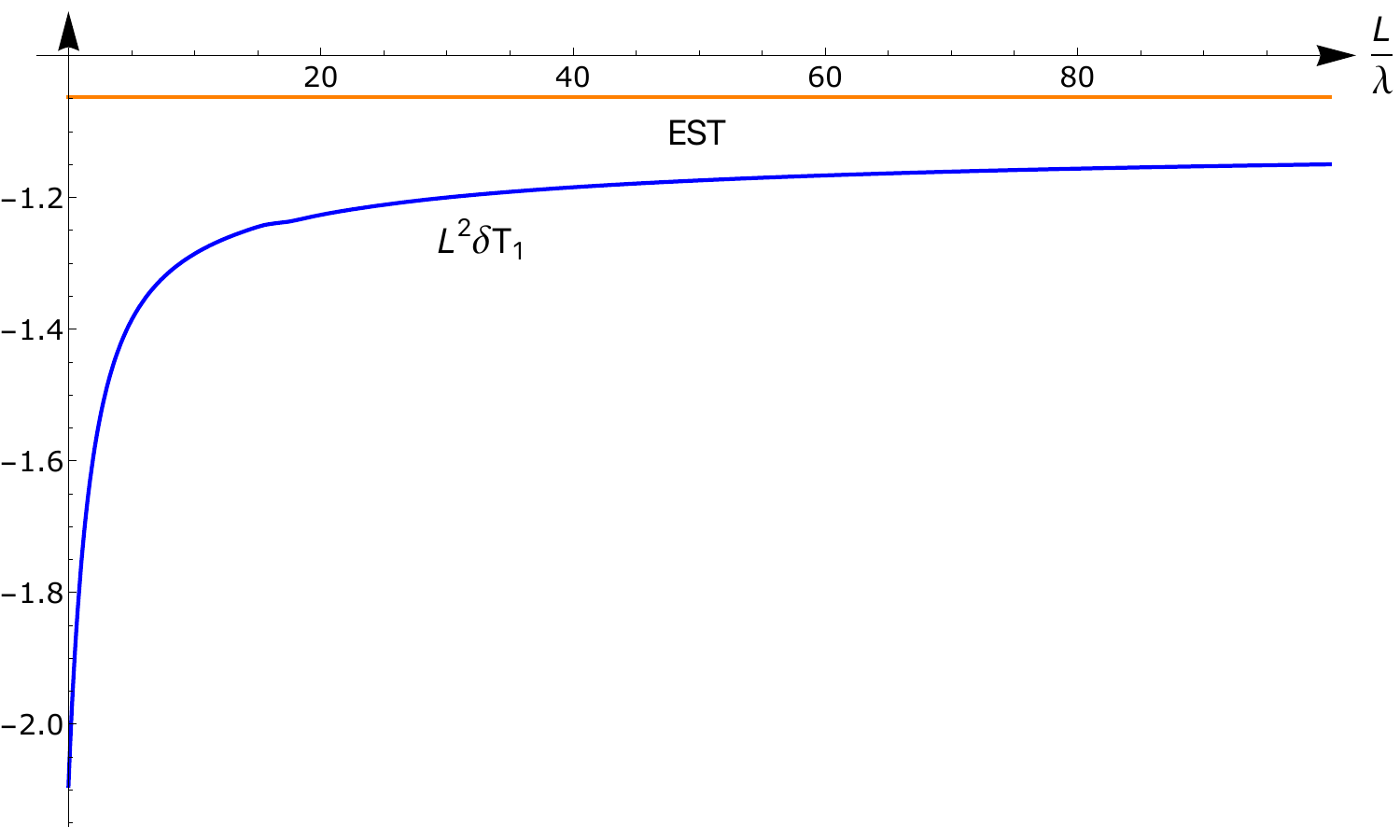}
 \caption{The contribution to the string tension at finite volume, compared to the EST prediction.}
   \label{fig:dTBulk}
\end{figure}

The intuition behind the slow decay of the function $J(x)$ is the following:  at low energies (i.e. $\kappa \ll 1$) the leading contribution to the phase shift comes from the angular mode $M=-1$ (or $s$-wave for the dual Hamiltonian) that vanishes as $1/\log(\kappa)$. All remaining partial waves vanish as power law $\delta_{M,n}\left(\frac{\lambda}{L}v\right) \approx c_{M,n} \left(\frac{\lambda}{L}v\right)^{\alpha}$ with $\alpha>0$ and their contribution to $J$ can be computed analytically
\begin{equation}
    \int_0 ^{\infty}d v\,\delta_{M,n}\left(\frac{\lambda}{L}v\right) v\sum_{r=1}^{\infty} K_{0}\left(rv\right) \sim c_{M,n}  \left(\frac{\lambda}{L }\right)^{\alpha}2^{\alpha}\Gamma\left(2+\frac{\alpha}{2}\right)^2 \zeta(\alpha+2)\,.
\end{equation}
The low energy regime of the phase shifts maps to the $\lambda\ll L$ limit of the Casimir energy, namely the long string regime, and the leading contribution comes from the $M=-1$ partial wave.

On the other hand, for the extreme regime $L/\lambda\to 0$ one obtains
\be
\lim_{x\to 0}J(x) = 1 \,.
\ee

\item For $n>1$, the modulus $\lambda$ becomes a dynamical degree of freedom, and the 
ground-state energy, together with its $L  \rightarrow\infty$ contribution, should be interpreted as an effective potential for this mode.

\end{itemize}
We can finally collect all the various contributions together to get the total quantum corrected ground-state energy of the fundamental string:
\be
E_{1}(L, g ,\lambda) = \frac{L}{g^2}\left(4\pi+ \frac{g^2}{\lambda^2}\left(\alpha +  \frac{4\pi}{3}y_1(\lambda^{-1}) \right)
    \right) -\frac{\pi}{3L}\left[1+J\left(L/\lambda\right)\right] \,.
\ee

We conclude this section by commenting on the interplay among the various scales appearing in the problem. On the one hand, there is $g$, which is the natural inverse cutoff for the bulk EFT and sets the classical tension. The dynamics of string objects within this theory introduce two additional length scales, namely the longitudinal extension $L$ and the intrinsic scale $\lambda$ which governs the intrinsic extension in the transverse directions. The validity of the results presented in this work requires the following hierarchies to hold
\be\label{eq: hierarchies}
L\gg g \quad , \quad \lambda \gg g \,.
\ee
The first inequality is required to tame the fluctuations of the localized NG bosons, hence rendering the effective theory on the string robust. The second inequality allows to study the interactions among the string and the bulk pions in a controlled way, hence being able to truncate the expansion at one-loop.

The important ratio $L/\lambda$ interpolates between two extreme regimes. On the one hand, at $L/\lambda\to \infty$ the string decouples from the bulk and its dynamics is accounted for by standard EST. On the other end, for small $L/\lambda$, the effects of the interactions dominate and the observables associated to the string depart considerably from the predictions of EST.

\section{Effective width of confining strings}\label{sec: 5}

An additional characteristic feature of confining strings is their effective width $\cW$, which measures the transverse extension of the flux tube. A quantitative definition of this quantity requires the choice of an observable sensitive to the spatial distribution of the flux in the directions orthogonal to the string. A natural candidate is the expectation value of the electric field component parallel to the string, $\langle E^\parallel(x_\perp)\rangle$, where $x_\perp$ denotes the coordinates transverse to the string profile.

In terms of this observable, the string width can be defined as a moment of the transverse electric field distribution. In the seminal work \cite{Luscher:1980iy}, this was taken to be the second moment. However, for certain field profiles this definition may suffer from divergences associated with heavy tails, \ie~with configurations whose decay at large transverse distances from the string is not sufficiently fast. In our specific case, we find more convenient to characterize the width through the first absolute moment of the distribution, defined as
\be\label{eq: mean square width}
\cW \;:=\; 
\frac{\int d^2 x_\perp \, |x_\perp| \, \langle E^\parallel(x_\perp)\rangle}
{\int d^2 x_\perp \, \langle E^\parallel(x_\perp)\rangle}\,,
\ee
which provides a physically meaningful measure of the transverse size of the flux tube.

In its crudest form, EST assumes an infinitely thin flux tube and provides a framework to quantify its effective width due to the quantum fluctuations of the worldsheet NG bosons. Naively, one would expect $\cW$ to be determined by a characteristic scale of the bulk theory, in particular the mass gap. This is referred to as the {\it intrinsic} width of the confining flux tube. In theories like Yang-Mills or QCD, such an intrinsic width should be inversely poportional to the strong coupling scale. However, even within the simple setup of EST, where only the dynamics of NG bosons is considered, the effective width turns out to be larger than expected and, moreover, tends to diverge logarithmically with the length $L$ of the flux tube
\be\label{eq: NG width}
\cW_{\text{EST}}^2 \propto \frac{D-2}{2\pi T}\log\left(\frac{L}{L_0}\right) \,,
\ee   
with some model dependent constant $L_0$ whose determination is typically beyond EST, and an order-one proportionality constant which depends on the specific definition of the width. Of course, for sufficiently long flux tubes, $L^{-1}\ll \sqrt{T}$, the profile still satisfies $\cW\ll L$ and the predictions of EST are robust. Using the definition \eqref{eq: mean square width}, together with the effective string theory prediction for the profile of the electric field \cite{Luscher:1980iy}, 
\begin{equation}
E_{\text{EST}}^\parallel(x_\perp)\propto\exp\!\left(-\frac{|x_\perp|^2}{\ell^2}\right)\,,
\end{equation}
where
\begin{equation}\label{eq: def of ell}
\ell^2 := \frac{\log(L/L_0)}{\pi T}\,,
\end{equation}
one readily obtains
\begin{equation}
\cW_{\text{EST}}
=\frac{\int d^2 x_\perp\, |x_\perp|\, e^{-|x_\perp|^2/\ell^2}}
{\int d^2 x_\perp\, e^{-|x_\perp|^2/\ell^2}}
=\frac{\sqrt{\pi}}{2}\,\ell\,.
\end{equation}

The logarithmic behavior in \eqref{eq: NG width} of the effective width is a natural consequence of the tendency of two dimensional NG bosons to spread out. Relatedly, the width is a sensible observable associated to open flux tubes between infinitely massive quarks separated by a finite distance. On the contrary, for closed strings this is not a well defined observable since the dynamics of worldsheet zero modes effectively delocalize the profile. In practice, however, an accurate estimate of the width is obtained by neglecting boundary effects and freezing the zero modes by hand. From this perspective, and as we will explicitly verify in our model, the width \eqref{eq: NG width} corresponds to the extension of the gaussian wavefunction associated to the NG bosons with their zero mode removed. 

On general grounds, the interplay between the massless gaussian wave function and the intrinsic width provides valuable physical information about the dynamics of confining strings. This simply stems from the fact that, from the perspective of the worldsheet, effects governed by the intrinsic width are naturally associated to non-universal interactions with bulk degrees of freedom. In models featuring the scale separation $M_{gap}\ll \sqrt{T}$, it is possible to quantify this interplay within a perturbative expansion \cite{Aharony:2024ctf}. 

In this section we consider a similar problem in the context of confining strings embedded in a gapless bulk. Logically, this scenario holds marked differences with respect to the case of $M_{gap}>0$. In particular, and considering the $n=1$ solution for concreteness, the role of the intrinsic width is played by the overall parameter $\lambda$. This quantity is intrinsic to the solution and is not determined from the effective field theory, hence from a bottom-up approach it may take arbitrary values. In this context, we will verify that quantum corrections dominate, enhancing the extension of the flux tube profile in the transverse directions with a width that depends logarithmically on $L$ as in \eqref{eq: NG width}.

\subsection{Stringy formalism}

An accurate determination of the width is not completely captured by standard perturbation theory, but rather requires a resumation of the effects originated by the massless fields on the worldsheet. This is achieved by a certain ``exponentiation'' of the NG bosons. This procedure was recently applied to compute the effective width of confining strings in three-dimensional QED \cite{Aharony:2024ctf}. Here, we briefly review the method and apply it to our case of interest, showing that considerations very similar to those of Ref.~\cite{Aharony:2024ctf} arise in the present context.

The first step is to promote the classical moduli associated to the broken translations to fields supported on the worldsheet, namely
\be\label{eq: NG fluct}
\omega(z,\sigma)=\omega_{cl}(z-z_0(\sigma))+\delta\omega(z-z_0(\sigma),\sigma) \,,
\ee
For notational simplicity, we use $z$ to collectively denote the two transverse coordinates $z$ and $\overline z$. Likewise, we employ $\sigma$ to denote collectively the coordinates along the string worldsheet. Throughout this discussion, we restrict our attention to a single pair of moduli, $(z_0,\overline z_0)$. As discussed in Section~\ref{sec: quadratic fluctuations}, for the fundamental flux tube ($n=1$) these are the only normalizable zero modes in the quantum theory. For higher--charge strings ($n>1$), one must instead account for the full set of $4n-2$ dynamical zero modes and their associated contributions to the quantum width.

Expanding around the classical solution, the dynamics of the collective coordinates $z_0(\sigma),\overline z_0(\sigma)$ arise from a simple implementation of the chain rule
\bea
&\partial_\mu \omega_{cl}\partial^\mu \overline \omega_{cl} + \partial_\mu \omega_{cl}\partial^\mu\delta \overline\omega + {\rm c.c.} +\partial_\mu \delta \omega\partial^{\mu} \delta\overline\omega = \\
&= \partial_z \omega_{cl}\partial_{\overline z} \overline \omega_{cl}\left[1+(\partial_\sigma z_0)^2\right] -\partial_\sigma z_0 \partial_z \omega_{cl} (\partial^\sigma - \partial^\sigma z_0\partial_z - \partial^\sigma \overline z_0\partial_{\overline z})\delta\overline \omega + \partial_{z}\omega_{cl}\partial_{\overline z}\delta \overline \omega + {\rm c.c.} \\
&\qquad \quad  + (\partial_\sigma - \partial_\sigma z_0\partial_z - \partial_\sigma \overline z_0\partial_{\overline z})\delta \omega(\partial^\sigma - \partial^\sigma z_0\partial_z - \partial^\sigma \overline z_0\partial_{\overline z})\delta\bar \omega +\partial_z \delta \omega\partial_{\overline z} \overline \omega
\eea
where, after acting with the derivatives, we can remove the $z_0$-dependence on the fields due to the fact that the transverse coordinates are integrated in the action. Therefore, one can derive an effective action involving the zero modes and their effective interactions with the bulk pion fluctuations. Notice that the fluctuations \eqref{eq: NG fluct} should be accompanied by bulk fluctuations, i.e. small fluctuations that depend on the full space–time coordinates, as discussed in Section~\ref{sec: quadratic fluctuations}. However, one can show that these give rise only to higher-order contributions to the width in the regime $g \ll \lambda$. We refer the reader to Appendix~\ref{app: stringy} for a comprehensive expansion of the effective action, including the leading interaction vertices. Here, we restrict ourselves to computing the lowest-order contribution to the string width.

The leading order quadratic piece involving the NG bosons reads
\be
S=\frac{2}{g^2}\int d^2\sigma d^2 z \frac{\partial_z \omega_{cl}\partial_{\overline z}\overline \omega_{cl}}{(1+|\omega_{cl}|^2)^2}\left(1+\partial_\sigma z_0\partial^\sigma \overline z_0\right) 
= T_{cl}+\int d^2\sigma \partial_\sigma Z_0\partial^\sigma \overline Z_0
\ee
where, in the last step, we integrated the classical profile to obtain the classical tension and we cannonically normalized the complex scalar fields
\be
Z_0=\sqrt{T_{cl}} \, z_0\,.
\ee
So far we have just presented an alternative expansion of the effective action, making it manifest the dependence on the NG bosons. The actual resummation comes about when computing the expectation value of a particular observable $\cO(z-z_0(\sigma))$. 
More precisely, we extract the dependence on the zero modes by means of the following manipulation
\bea
\langle \cO(z-z_0(\sigma))\rangle &= \int d^2\tilde z \langle \cO(z-\tilde z) \delta^{(2)}(\tilde z-z_0(\sigma))\rangle \\
&=\int d^2\tilde z \int\frac{d^2p}{(2\pi)^2}e^{i p\cdot \tilde z}\langle e^{-ip\cdot z_0(\sigma)}\cO(z-\tilde z)\rangle
\eea
where in the last step we used the Fourier representation of the delta functional in terms of a complex two-momenta $p=(p,\overline p)$ and 
\be
p\cdot z_0(\sigma) = \overline p z_0(\sigma) + p \overline z_0(\sigma)\,.
\ee 

Now, consider the case in which the expectation value of the observable in question has a leading contribution given by the classical string profile
\be
\cO(z-\tilde z)=\cO_{cl}(z-\tilde z) + \ldots
\ee
where higher order contributions are suppressed by higher powers of the inverse cutoff $g\sim \Lambda^{-1}$. 
At this order, the dependence on the NG bosons occurs only through the vertex operator, hence obtaining
\be
\langle \cO(z-z(\sigma))\rangle  = \int d^2\tilde z \int\frac{d^2p}{(2\pi)^2} e^{i p\cdot \tilde z}\langle e^{-ip\cdot z_0(0)}\rangle \cO_{cl}(z-\tilde z)
\ee
where we used the preserved translation symmetry along the worldsheet directions to fix $\sigma=0$.

We therefore see that the leading order contribution is determined by the expectation value of a vertex operator within the theory of a free complex scalar boson. A standard computation then yields
\be\label{eq: conv classical}
\langle \cO(z-z(0))\rangle = \frac{T_{cl}}{2\pi G(0)}\int d^2 \tilde z e^{-\frac{T_{cl}}{2 G(0)}|\tilde z|^2} \cO_{cl}(z-\tilde z) + o(g^2)\,.
\ee 

A few remarks are in order. First, the expectation value of a single vertex operator for a closed string vanishes due to the shift symmetry of the theory, as the integration over the zero mode imposes charge conservation. By this token, the mean width is not a well defined quantity for closed strings, but rather for open strings. To the order we are working, and for long enough open strings, boundary effects can be neglected. Therefore, in practice, one can just remove the zero mode in order to obtain a consistent result \cite{Aharony:2024ctf}. 

Second, the expressions obtained above are just formal, due to the fact that we are evaluating the two dimensional Green's function at coincident points. In order to get a finite result one needs to regularize this quantity. Following the same dimensional regularization procedure as in \cite{Aharony:2024ctf} we get\footnote{The relative factor of 2 arises because we are dealing with a complex scalar.} 
\be
G(0)=-\frac{1}{\pi \epsilon}+\frac{2\log(\mu L)+\gamma_E-\log(4\pi) }{2\pi}+\cO(\epsilon)\,.
\ee
In the expression above, we can trade the sliding scale $\mu$ by a reference scale $M$ that comes into the definition of the regulator
\be
G(0)=-\frac{1}{\pi \epsilon_M}+\frac{\log(M L)}{\pi}+\cO(\epsilon)\,.
\ee
The extracted divergence can be used to rewrite vertex operators in terms of normal ordered ones
\bea\label{eq: noral order}
e^{-i\frac{p}{\sqrt{T}}\cdot Z_0}&=e^{-i\frac{p}{\sqrt{T}}\cdot Z_0} e^{-\frac{p\overline p}{\pi T \epsilon_M}+\frac{p\overline p}{\pi  T \epsilon_M}} \\
&= :e^{-i\frac{p}{\sqrt{T}}\cdot Z_0}: \left( 1 +\frac{p\overline p}{\pi T \epsilon_M} +\ldots \right)\,.
\eea
Expectation values for the properly normal ordered vertex operators are finite by construction and depend on the reference renormalization scale $M$. Within the perturbative expansion, divergences in \eqref{eq: noral order} arise as higher loop effects and should be properly cancelled by higher order divergent diagrams. An explicit check of the divergence cancellation at one loop, as it was done in \cite{Aharony:2024ctf} for $QED_3$, is technically tedious and will not be presented in this article. 
On general grounds, the a priori arbitrary scale $M$ will contribute to higher loops through logarithms of the form $\log(\cW_{int} M)$, with $\cW_{int}$ some intrinsic length scale associated to the solitonic string. This a common feature in effective field theory, as opposed to renormalizable UV complete theories, and a sensible choice of the physical cutoff is required in order to avoid logarithmic enhancements that may invalidate the perturbative expansion.  For the case at hand, the only natural possibility is for the intrinsic width to be proportional to the overall size of the classical configuration $\lambda$ hence implying $M\sim \lambda^{-1}$ up to an $\cO(1)$ proportionality constant. We will henceforth adopt this choice, obtaining the finite leading order contribution 
\be\label{eq: ren conv classical}
\langle \cO(z-z(0))\rangle \approx \frac{T_{cl}}{2\pi G_\lambda(0)}\int d^2 \tilde z e^{-\frac{T_{cl}}{2 G_\lambda(0)}|\tilde z|^2} \cO_{cl}(z-\tilde z)
\ee
with
\be
G_\lambda(0) =\frac{1}{\pi} \log\left(\frac{L}{\lambda}\right)\,.
\ee

\subsection{Effective width of the fundamental string}

Given the general result \eqref{eq: ren conv classical}, we can now specialize to our case of interest. For theories with a $U(1)$ 1-form symmetry, the electric field can be defined through the conserved current as
\begin{equation}
    E_i = J_{0i} \, ,
\end{equation}
so that 
\be\label{eq: electric field n=1}
E_{cl}^\parallel(z-\tilde z)=  \frac{\partial \omega_{cl}{\overline\partial\overline\omega_{cl}}}{\left(1+|\omega_{cl}|^2\right)^2} = \frac{\lambda^2}{\left(|z-\tilde z|^{2}+\lambda^{2}\right)^2}\,.
\ee
Notice that, because of the relation \eqref{eq: BP bound}, this coincide with the energy density operator.

Combining all the ingredients together, we can express the width at order $g^0$ as
\begin{equation}\label{eq: width n=1}
    \cW(\lambda,\ell)\equiv \frac{\int d^2z |z|\langle E^\parallel(z)\rangle}{\int d^2z \langle E^\parallel(z)\rangle}=\frac{\int d^2z d^2\tilde z\; |z|e^{-\frac{|\tilde z|^2}{\ell^2}} E_{cl}^\parallel(z-\tilde z)}{\int d^2z d^2\tilde z\; e^{-\frac{|\tilde z|^2}{\ell^2}}E_{cl}^\parallel(z-\tilde z)}\,.
\end{equation}
Plugging \eqref{eq: electric field n=1} into \eqref{eq: width n=1} and doing some simple change of variables, we get
\bea\label{eq: width integral}
\cW(\alpha,\ell)= \ell \frac{8}{\pi} \int_0^\infty dx dy \,\frac{y^4 x(1+\alpha x)}{(1+y^2 x^2)^2}\mathbb{E}\left(\frac{4\alpha x}{(1+\alpha x)^2}\right)e^{-y^2}
\eea
where $\ell$ is defined in \eqref{eq: def of ell}, $\alpha := \frac{\lambda}{\ell}$ and the standard elliptic integral is defined as
\be
\mathbb{E}(x)=\int_0^{\frac{\pi}{2}}d\theta \sqrt{1-x\sin^2\theta}\,.
\ee
We notice that in the limit $\alpha \rightarrow 0$ we get
\be
\cW(\alpha=0,\ell)\equiv \cW_{EST}(\ell)\,.
\ee
Again we conclude that in this limit the string decouples from the bulk dynamics and the EST prediction is recovered.

We can now compute the deviation from the EST, by computing the $\alpha \ll 1$ corrections. When $\alpha \not=0$ it is useful to split the integral over $x$ in the two regions $x\alpha\gtrless 1 $. When $x\alpha < 1$ we can safely expand the integrand, obtaining
\bea
\frac{8\ell}{\pi} \int_0^{\infty} dy \int_0^{\alpha^{-1}}dx\frac{y^4x e^{-y^2}}{(1+x^2 y^2)^2}\frac{\pi}{2}\left(1-\frac{\alpha^2 x^2}{4}+\ldots\right)=\frac{\sqrt{\pi}\ell}{2}+\frac{\sqrt{\pi}\ell}{4}\alpha^2\left(\log\left(\frac{1}{4\alpha^2}\right) -5-\gamma_E\right)\,.
\eea 
For $x \alpha >1$, we get
\bea
\frac{8\ell}{\pi}
\int_0^{\infty} dy \int_{\alpha^{-1}}^\infty dx\,
\frac{y^4 x (1+\alpha x) e^{-y^2}}{(1+x^2 y^2)^2}
\mathbb{E}\!\left(\frac{4\alpha x}{(1+\alpha x)^2}\right)
\\
= \frac{8\ell}{\pi}\alpha^2
\int_0^{\infty} dy \int_1^\infty dx\,
\frac{y^4 x (1+x) e^{-y^2}}{(\alpha^2+x^2 y^2)^2}
\mathbb{E}\!\left(\frac{4x}{(1+x)^2}\right)\,.
\eea
Again, expanding for small $\alpha$ we get
\be
\frac{8\ell}{\pi} \alpha^2\int_0^{\infty} dy e^{-y^2} \int_1^\infty dx\frac{1+ x}{x^3}\mathbb{E}\left(\frac{4 x}{(1+ x)^2}\right)
\ee
where\footnote{In the result below $G\approx 0.915$ is the Catalan constant. We thank Julio Parra-Martinez for pointing out that the integral below can be performed analytically.}
\be
\int_1^\infty dx\frac{1+ x}{x^3}\mathbb{E}\left(\frac{4 x}{(1+ x)^2}\right)=\frac{5}{4}+ \frac{G}{2} \approx 1.7079\cdots
\ee
Combining everything together, we get the small $\alpha$ expansion of the width
\be
\cW(\alpha,\ell)=\frac{\sqrt{\pi}}{2}\ell+ \alpha^2\ell(C - \frac{\sqrt{\pi}}{2}\log(2\alpha)) + o(\alpha^3)
\ee
with $C\approx 1.3831$. 
We notice that finite $\alpha$ contributions tend to increase the string width with respect to the EST value. It is also worth noticing that the $\alpha\ll 1$ regime is verified for exponentially large strings, namely
\be
L\gg \lambda \, e^{\lambda/g}
\ee
since we still demand $\lambda\gg g$ in order to be within the regime if validity of the EFT (see discussion around \eqref{eq: hierarchies}). This behavior is analogous to the one evidenced by confining strings in three-dimensional QED \cite{Aharony:2024ctf}.

\begin{figure}
    \centering
    \includegraphics[scale=0.8]{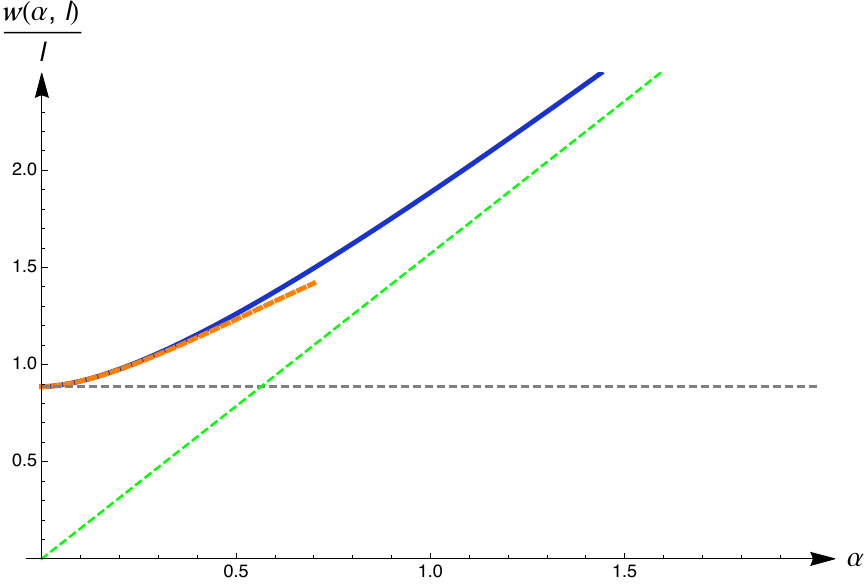}
     \caption{String width as a function of the dimensionless parameter $\alpha=\lambda/\ell$. 
The full numerical result obtained from \eqref{eq: width integral} is shown by the solid blue curve. 
The dashed orange (green) curve represents the analytical approximation valid in the regime $\alpha\ll 1$ ($\alpha\gg 1$), while the dashed gray line corresponds to the EST prediction.}

    \label{fig: width}
\end{figure}

Finally, it is instructive to study how the mean width $\cW(\alpha,\ell)$ asymptotes for large values of $\alpha$. A straightforward computation shows that in this regime
\be
\cW(\alpha,\ell)\simeq \frac{\pi}{2}\lambda + O(\alpha^{-1}) 
\ee
which is precisely the mean width given purely by the classical solution, that is without taking the quantum expectation value for the electric field
\be
\cW_{cl}=\frac{\int d^2z |z| E^\parallel(z)}{\int d^2z  E^\parallel(z)}=\frac{\pi}{2}\lambda
\ee
We conclude that in the $\alpha=\lambda/\ell\to \infty$ limit, the fluctuation of the NG bosons become negligible, consistently with the fact that their gaussian wave function gets sharply peaked at the origin in the transverse plane. 

In Figure \ref{fig: width} we compare the exact result obtained from \eqref{eq: width integral} with the various analytical approximations for $\alpha\ll1$ and $\alpha\gg1$.

\section{Comments on UV completions and stability of confining strings}\label{sec: 6}

While the philosophy of this work is to study the features of confined string-like objects from an EFT point of view, it is interesting to comment on the interpretation of these strings when the $\bC\bP^1$ model is regarded as the IR fixed point of RG flows triggered by certain UV completions.

\paragraph{Abelian--Higgs Model.} One of the simplest UV completions of the $\bC\bP^1$ model is the Abelian--Higgs model (AHM) with $\cN_f=2$ charge-one complex scalars, defined by the action
\begin{equation}
S_{\text{AHM}} =\int d^4 x \left(-\frac{1}{4e^2}F_{\mu \nu}F^{\mu\nu} + \sum_{i=1,2}|D_{\mu}\phi_i|^2 - m^2 |\phi_i|^2 - \lambda_\phi |\phi_i|^4\right)\,,
\end{equation}  
When $m^2<0$ the scalar dublet $\phi_i$ condenses and the $SU(2)$ flavor symmetry is spontaneously broken to its $U(1)$ subgroup leading to a $\bC\bP^1$ phase.

This theory enjoys a $U(1)$ magnetic one-form symmetry acting on ’t~Hooft lines, which is naturally matched by the $U(1)$ topological symmetry of the NLSM. It is well known that the Abelian Higgs model admits dynamical and stable string excitations, namely Abrikosov--Nielsen--Olesen (ANO) vortices \cite{Abrikosov:1956sx,NIELSEN197345}\footnote{See also the recent discussion in \cite{Dumitrescu:2025fme} which is more in line with the discussion presented in this paper.}, as well as non-stable string-like configurations known as semilocal strings \cite{Vachaspati:1991dz,Hindmarsh:1992yy,LEESE1993639}. The latter are characterized by a size modulus $\lambda$ and reduce to the ANO vortices in the $\lambda \to 0$ limit. Although semilocal strings are generically unstable, they become effectively stable in the IR limit, where they interpolate to the $\bC\bP^1$ strings studied in this work. We therefore conclude that the family of infrared string configurations parametrized by $\lambda$ is not stable at high energies: ultraviolet completions generically lift this modulus and select particular values of $\lambda$ corresponding to stable string solutions at all energy scales. In the case of AHM, such value is $\lambda = 0$ which is singular in the IR EFT: for any arbitrarily small but nonzero value of $\lambda$, the classical configuration \eqref{eq: charge n rot sym} carries tension $T_n$ and one-form symmetry charge $n$, whereas at $\lambda = 0$ the profile collapses to the vacuum and no string is present. This pathology is however resolved by the UV completion. From the perspective of the low energy $\bC\bP^1$ model, the modulus $\lambda$ gets destabilized by including higher derivative corrections, which explicitly break the emergent scale invariance to which $\lambda$ is naturally associated.  

\paragraph{Adjoint QCD.}A more interesting UV theory flowing to the $\bC\bP^1$ phase is $SU(N_c)$ gauge theory with $N_f=2$ adjoint Weyl fermions, described by the action
\be\label{eq: Adj QCD action}
S_{\text{AdjQCD}}=-\int d^4x \left(\frac{1}{4g_{YM}^2}{\rm Tr}\left[F_{\mu\nu}F^{\mu\nu}\right] + i\sum_{j=1,2}{\rm Tr}\left[\bar q_j\bar\sigma^\mu D_\mu q_j\right]\right)\,,
\ee
where the trace is performed over color indices, $\bar\sigma^\mu=(-\mathbb{I},\sigma^1,\sigma^2,\sigma^3)$ with $\sigma^a$ the standard Pauli matrices and $D_\mu$ denotes the covariant derivative in the adjoint representation of the gauge group. It is conjecturally expected that this theory confines and breaks spontaneously the chiral symmetries via the the condensation of the fermion bilinear $ \Tr(q_{(i}q_{j)})$. The symmetry breaking pattern is
\begin{equation}
    \frac{SU(2)\times \bZ_{4N_c}}{\bZ_2}\;\;\longrightarrow \;\;U(1)\rtimes \bZ_2 \,,
\end{equation}
leading to $N_c$ degenerate gapless vacua described by a $\bC\bP^1$ NLSM (see \cite{Cordova:2018acb} for more details).\footnote{For the purposes of our discussion we can neglect the $N_c$ degeneracy and focus on a single $\bC\bP^1$ theory. This is formally correct if we consider the theory on infinite volume.} While this scenario is conjectural, extrapolation from soft supersymmetry breaking gives strong evidence in its favor \cite{Cordova:2018acb,DHoker:2024vii}.
 
Based on the AHM discussion, it is tempting to conclude that for some values of $\lambda$ the solitonic strings of the NLSM interpolate to QCD strings in the UV description. While this can be true for small values of $N_c$ (e.g. this is likely true for $N_c=2$, see \cite{Cordova:2018acb}, where the $\bZ_2$ UV 1-form symmetry is canonically embedded in the U$(1)$ IR 1-form symmetry) we are going to show that the same conclusion cannot be true at large $N_c$.\footnote{We thank G. Cuomo for discussions on this point.} This can be argued by looking at the large $N_c$ behavior expected for the sigma model coupling $g$. Standard large-$N_c$ arguments show that
\begin{equation}
    g^{-1} \sim \Lambda_{\text{QCD}}N_c \;\;\Longrightarrow\;\; S_{\bC\bP^1}\sim \Lambda_{\text{QCD}}^2N_c^2\,.
\end{equation}
However, this behavior implies that the tension of the solitonic strings diverges in the large--$N_c$ limit as $T \sim N_c^2\,\Lambda_{\text{QCD}}^2$. This scaling does not comply with expectations for confining strings in large-$N_c$ QCD, where one expects the existence of stable flux tubes with a tension set by the strong coupling scale, $T \sim \Lambda_{\text{QCD}}^2$, and independent of $N_c$ at leading order. 

This mismatch between $\bC\bP^1$ strings and adjoint QCD can be understood from the abelian phase described in \cite{Cordova:2018acb}. At intermediate energies, the (supersymmetric) adjoint QCD reduces to a $U(1)^{N_c-1}$ Abelian gauge theory, which is further Higgsed to the $\bC\bP^1$ NLSM. For $N_c=2$, the Abelian theory has a single $U(1)$ 1-form symmetry, so the identification with the IR 1-form symmetry is immediate. For $N_c>2$, however, the gauge theory possesses a larger 1-form symmetry, and the $\bC\bP^1$ strings do not directly match the UV Wilson lines.

A natural question is whether the behavior of the heavy solitonic strings studied in this work—such as their ground state energy and width—extends, at least qualitatively, to more fundamental UV flux tubes in adjoint QCD, which remain embedded in a gapless bulk pion theory. Addressing this issue, however, lies beyond the scope of the present article.

\section*{Acknowledgments}

We are grateful to Riccardo Argurio, Gabriel Cuomo, Victor Gorbenko, Guzman Hernandez-Chifflet, Marco Meineri, Julio Parra-Martinez and Alessandro Podo for useful feedback.  The research of G.G.~is funded through an ARC advanced project, and further supported by IISN-Belgium (convention 4.4503.15). The work of J.A.D. is funded by the Spanish MCIN/AEI/10.13039/501100011033 grant PID2022-126224NB-C21.

\appendix

\section{Bulk renormalization and consistency of the effective string action}\label{app:betaf}

\subsection{Counterterms for four-derivative couplings }

In Section \ref{sec: 4} we emphasized the importance of fixing a particular renormalization scheme within the four-dimensional theory in order to obtain finite results for the observables associated to the solitonic objects studied in this work. 

In particular, at one-loop, it is required to include the bulk four-derivative couplings \eqref{eq: 4 derivative S}, that we report here again for convenience 
\be\label{eq: app 4 derivative S}
S_{4\partial} =\int d^4 x \left[  y_1 \frac{(\partial_\mu \omega\partial^\mu\bar\omega)^2}{\left(1+|\omega|^2\right)^4}+ y_2  \frac{(\partial_\mu \omega\partial^\mu\omega)(\partial_\mu \bar\omega\partial^\mu\bar\omega)}{\left(1+|\omega|^2\right)^4}\right] \,,
\ee
with the dimensionless bare couplings $y_i=y_i^r+\delta y_i$ including a renormalized piece and a divergent counterterm $\delta y_i$. At the order we are interested in, the densities multiplying these couplings can be treated classically, since their contribution to interaction vertices are further suppressed by higher orders in $g$. 

As explained in the main text, the leading divergencies and scale dependence of $\delta y_i$ and $y_i^r$ respectively are already fixed by the bulk dynamics. Here we provide a derivation of these by means of the Heat Kernel expansion \cite{Vassilevich:2003xt}. In particular, for a given second order differential operator $P$ one defines
\begin{equation}
    K(t)= \text{Tr}\left(e^{-tP}-e^{-tP_0}\right)\, , 
\end{equation}
with $P_0=-\partial_\mu\partial^\mu$. The Heat Kernel $K(t)$ admits the following  asymptotic expansion for small $t$
\begin{equation}
    K(t)= \frac{1}{t^{d/2}}\sum_{n=0}^{\infty}t^{n}a_{2n}\, ,
\end{equation}
where the coefficients $a_{2n}$ are integrals of local functions determined by $P$. 

With this at hand, the regularized effective action can be written as\footnote{Note that the different choice of regulator with respect to the main text only affects the finite part of the renormalized couplings, hence being inconsequential for our purposes.}
\begin{equation}
    W(s)= -\frac{1}{2}\mu^{2s}\int_0^{\infty}\frac{dt}{t^{1-s}}K(t)\, , 
\end{equation}
whose logarithmic divergence in the limit $s\rightarrow 0$ is captured by the heat kernel coefficient $a_d$
\begin{equation}
    W=-\frac{1}{2}\left(\frac{1}{s}-\gamma_E+\log(\mu^2)\right)a_d + \text{finite}\, .
\end{equation}
For our four-dimensional differential operator $P$ we find (see \eqref{eq: X} and $\eqref{eq: F}$ for definitions)
\begin{equation}
\begin{split}
     a_4&=\frac{1}{(4\pi)^2}\int d^4 x \left(\frac{1}{2}\text{Tr}(X^2)+ \frac{1}{12}\text{Tr}(F_{\mu \nu}F^{\mu \nu})\right)\\&= \frac{1}{6\pi^2}\int d^4 x \cL^{(1)}+  \frac{1}{3\pi^2}\int d^4 x \cL^{(2)}\, ,
\end{split}
\end{equation}
hence, within a minimal subtraction scheme, the above computation yields
\begin{equation}\label{eq: ct}
    \delta y_1= -\frac{1}{12\pi^2}\left(\frac{1}{s}+\gamma_E\right)\, , \qquad \delta y_2= -\frac{1}{6\pi^2}\left(\frac{1}{s}+\gamma_E\right)\, ,
\end{equation}
and correspondingly 
\be
\beta_{y_1}=\frac{1}{6\pi^2} \quad , \quad \beta_{y_2}=\frac{1}{3\pi^2}\,.
\ee

\subsection{Trace Formula for $\beta$}

Here we provide an alternative determination of the scheme independent coefficient $\beta$, without resorting to the phase shift. As emphasized in the main text, this also stands as a check of consistency among the effective string theory and the bulk four dimensional effective action. In particular, we will find that the value of of $\beta$ is perfectly consistent with the beta functions computed in the previous subsection.

Again, the idea is to use the Heat Kernel expansion method, now applied to the two-dimensional operator describing transverse fluctuations around the string
\begin{equation}
    \widetilde{P}_n= -\left(D_a D^a +2\frac{\partial_a \omega_n\partial^a \omega_n}{(1+|\omega_n|^2)^2}\right)\unit\, ,
\end{equation}
to obtain some "trace formulas". In particular, assuming an axially symmetric string configuration and subtracting the vacuum contribution, we can write the Heat kernel as
\begin{equation}
    K(t)= -\frac{2}{\pi}\int_0^{\infty}d\kappa e^{-t\kappa^2/\lambda^2}\frac{d\delta_n(\kappa)}{d\kappa}\, . 
\end{equation}
where the factor of $2$ comes from the trace over the internal space. We then identify the coefficients of the small $t$ expansion with the moments of the density of states, in particular the coefficient of the term linear in $t$ is proportional to $\beta$. On the other hand, since $\widetilde{P}_n$ is two-dimensional, the small $t$ expansion reads 
\begin{equation}
    K(t)=\frac{1}{t}\sum_{n=0}^{\infty}t^{n}a^{(2)}_{2n}\, ,
\end{equation}
therefore leading o the following relation among $\beta$ and the coefficient $a^{(2)}_4$
\begin{equation}
    \frac{2}{\pi \lambda^2}\int_0^{\infty}d\kappa \frac{d\delta}{d\kappa} \kappa^2 = a^{(2)}_4\, \Rightarrow \beta = \frac{a_4^{(2)}\lambda^2}{8\pi}\, .
\end{equation}
Now, for the operator $\widetilde{P}_n$ we have
\begin{equation}
    a^{(2)}_4=\frac{1}{8\pi}\int d^2x \Tr\left[\left(2\frac{\partial_a\omega_n\partial^a\overline{\omega}_a}{(1+|\omega_a|^2)^2} \right)^2\right] + \frac{1}{48\pi}\int d^2x \Tr\left[F_{ab}F^{ab}\right]\, ,
\end{equation}
which, given
\begin{equation}
    F_{ab}= \frac{2}{(1+|\omega|^2)^2}\left(\partial_a\omega \partial_b\overline{\omega}-\partial_b\omega \partial_a\overline{\omega}\right) \, ,
\end{equation}
evaluates to 
\begin{equation}
   a^{(2)}_4=\frac{1}{\pi}\frac{2}{3}\int d^2x \left(\frac{\partial_a\omega_n\partial^a\overline{\omega}_n}{(1+|\omega_a|^2)^2} \right)^2=\frac{2}{3 \pi} \lambda^2B_n \, ,
\end{equation}
where $B_n$ is the same number that shows up in the counterterm contribution to the string tension in \eqref{eq: tension counterterm}. This then gives
\begin{equation}
    \beta = \frac{1}{12\pi^2}B_n\, ,
\end{equation}
or, for $n=1$
\begin{equation}
    \beta=\frac{1}{9\pi}\, .
\end{equation}

\section{Phase shift integrals}\label{app:PSints}
In this appendix we compute integrals involving the total phase shift $\delta_1(\kappa)$ using both numerical and analytical methods. We start with the simple integral
\begin{equation}
    I_0= \int_0^{\infty} d\kappa\,  (\delta_1(\kappa)-\pi)\, , 
\end{equation}
which enters in the Casimir energy of Skyrmions in $2+1$ dimensions. Writing
\begin{equation}
    \delta_1(\kappa)= R_{{l_{max}}}(\kappa)+ \widetilde{\delta}_1(\kappa)\, , 
\end{equation}
we have
\begin{equation}
    I_0= \int_0^{\infty}d\kappa \, R_{{l_{max}}}(\kappa) + \int_0^{\infty} d\kappa\,  (\widetilde{\delta}_1(\kappa)-\pi)\, . 
\end{equation}
The analytical part is simple to compute using the integral representation 
\begin{equation}
    \widetilde{\delta}_1(\kappa)= \pi + 4\pi^2\frac{d}{d\kappa}\int_0^{\infty}dt \cosh(t) \frac{\kappa^2 e^{2 \pi \kappa \cosh(t)}}{(e^{2 \pi \kappa \cosh(t)}-1)^2}\, ,
\end{equation}
we find
\begin{equation}
     \int_0^{\infty} d\kappa\,  (\widetilde{\delta}_1(\kappa)-\pi)= 4\pi^2\int_0^{\infty}dt \cosh(t) \frac{\kappa^2 e^{2 \pi \kappa \cosh(t)}}{(e^{2 \pi \kappa \cosh(t)}-1)^2}\Bigg|_{\kappa=0}^{\kappa=\infty}\, , 
\end{equation}
and using 
\begin{equation}
    \int_0^{\infty}dt \cosh(t) \frac{\kappa^2 e^{2 \pi \kappa \cosh(t)}}{(e^{2 \pi \kappa \cosh(t)}-1)^2}\simeq \frac{1}{4\pi^2}\int_0^{\infty}dt \frac{1}{\cosh(t)} +O(\kappa^2)=\frac{1}{8\pi}+O(\kappa^2)\, ,
\end{equation}
we have
\begin{equation}
    \int_0^{\infty} d\kappa\,  (\widetilde{\delta}_1(\kappa)-\pi)= -\frac{\pi}{2}\, . 
\end{equation}
It remains to evaluate the contribution of the numerical part $R_{{l_{max}}}(\kappa)$, so let us define
\begin{equation}
    I^{(0)}_{num}({l_{max}}, \kappa_{\text{max}}, \kappa_{\text{min}})=\int_{\kappa_{\text{min}}}^{\kappa_{\text{max}}}d\kappa R_{l_{max}}(\kappa)\, ,
\end{equation}
clearly the exact value is recovered for ${l_{max}}, \kappa_{\text{max}}\rightarrow \infty$ and $\kappa_{\text{min}}\rightarrow 0$. We will proceed by fixing $\kappa_{\text{max}}$ and $\kappa_{\text{min}}$ to some reasonable values and checking how the result is converging with ${l_{max}}$, see Table\ref{tab:Ints} for the results. Improving the extrema of integrations does not significantly modify the numbers reported.  
Our best estimate then is
\begin{equation}
    I_0= -\frac{\pi}{2}+I^{(0)}_{num}(70,10, 10^{-4})=-1.607 \,\, ,
\end{equation}
which is compatible with the results of \cite{Moss:1999xs,PhysRevB.61.2819}. 
Another integral we want to compute is the one that determines $\beta$
\begin{equation}
    \beta= \frac{1}{4\pi^2}\int_0^{\infty}d\kappa\, \kappa^2 \frac{d \delta_1}{d\kappa}\, ,
\end{equation}
which we rewrite integrating by parts 
\begin{equation}
    \beta= -\frac{1}{2\pi^2}\int_0^{\infty}d\kappa\, \kappa (\delta_1-\pi)\, ,
\end{equation}
and separating again the analytic part from the numerical one
\begin{equation}
    \beta= -\frac{1}{2\pi^2}\int_0^{\infty}d\kappa\, \kappa (\widetilde{\delta}_1-\pi)-\frac{1}{2\pi^2}\int_0^{\infty}d\kappa\,\kappa  R_{l_{max}}(\kappa)\, .
\end{equation}
The analytic part is simple to compute
\begin{equation}
    -\frac{1}{2\pi^2}\int_0^{\infty}d\kappa\, \kappa (\widetilde{\delta}_1-\pi)=2\int_0^{\infty}d\kappa \int_0^{\infty}dt \cosh(t) \frac{\kappa^2 e^{2 \pi \kappa \cosh(t)}}{(e^{2 \pi \kappa \cosh(t)}-1)^2}=\frac{1}{12\pi}\, ,
\end{equation}
while for the numerical part we define
\begin{equation}
    I^{(1)}_{num}({l_{max}}, \kappa_{\text{max}}, \kappa_{\text{min}})=\int_{\kappa_{\text{min}}}^{\kappa_{\text{max}}}d\kappa \, \kappa R_{l_{max}}(\kappa), 
\end{equation}
also here we fix the integration limits and check how the integral converges with ${l_{max}}$, our results are reported in Tab.\ref{tab:Ints}. Therefore our best estimate for $\beta$ is
\begin{equation}
    \beta= \frac{1}{12\pi}-\frac{1}{2\pi^2}I_{num}^{(1)}(70,10, 10^{-4})=0.035
\end{equation}
which is in great agreement with the analytical result obtained in Appendix \ref{app:betaf}, i.e.
\begin{equation}
    \beta= \frac{1}{9\pi}=0.035\cdots\, .
\end{equation}

\begin{figure}
    \centering
    \includegraphics[scale=0.4]{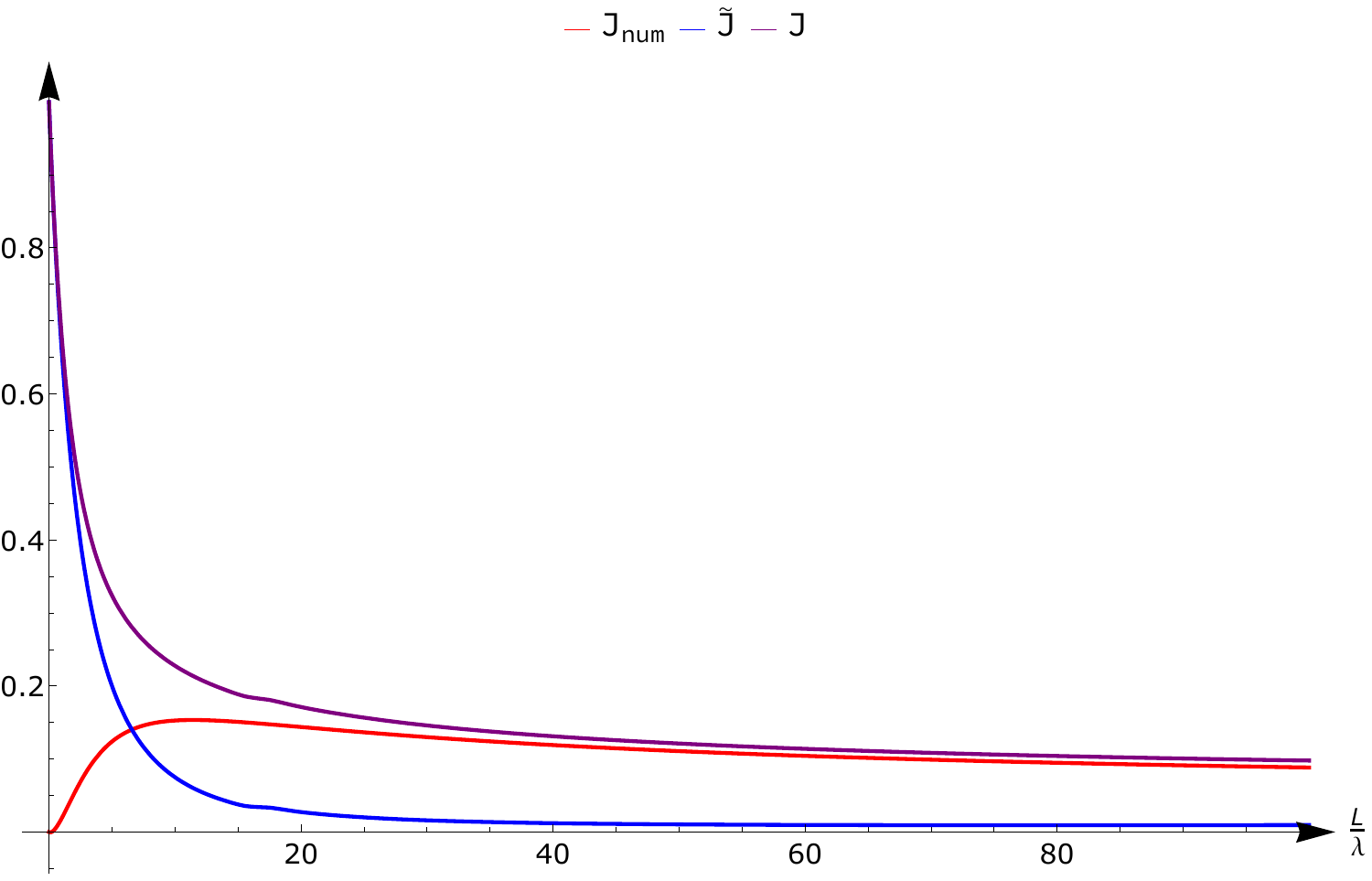}
    \caption{Plot of $J_{num}$, $\widetilde{J}$ and their sum $J$ as a function of $L/\lambda$.}
    \label{fig:dTApp}
\end{figure}

\begin{table}[h!]
\centering
\begin{tabular}{|c|c|c|c|c|} 
 \hline
 ${l_{max}}$& $10$ & $20$ & $50$ & $70$ \\ 
 \hline
$I^{(0)}_{num}$ & $-0.052$ & $-0.044$ & $-0.037$ & $-0.036$ \\ 
 \hline
$I^{(1)}_{num}$ & $-0.226$ & $-0.221$ &$ -0.177$  & $-0.174$ \\ 
 \hline
\end{tabular}
\caption{Values of the numerical integrals for $\kappa_{\text{min}}=10^{-4}$ and $\kappa_{\text{max}}=10$.}
\label{tab:Ints}
\end{table}
Another relevant integral to compute is the one determining the tension of the fundamental string in finite volume:
\begin{equation}
   J(z)\equiv -\frac{3}{\pi} L ^2\delta T_{1}(v) =z^{2}\frac{6}{\pi^3}\int_0^{\infty}d\kappa\, \int_0^{\infty}dt\, \delta_1(\kappa) \frac{\kappa}{e^{\kappa z\cosh{t}}-1}
\end{equation}
where we defined $z= L/\lambda$. As usual we separate the analytic from the numerical part in the phase shift
\begin{equation}
 \begin{split}
      J(z)=& z^{2} \frac{6}{\pi^3}\int_0^{\infty}d\kappa\, \int_0^{\infty}dt R_{l_{max}}(\kappa)\frac{\kappa}{e^{\kappa z\cosh{t}}-1} + z^2\frac{6}{\pi^2}\int_0^{\infty}d\kappa\, \int_0^{\infty}dt\frac{\kappa}{e^{\kappa z\cosh{t}}-1}\\ &+ z^{2}\frac{6}{\pi^3}\int_0^{\infty}d\kappa\, \int_0^{\infty}dt (\widetilde{\delta}_1(\kappa)-\pi)\frac{\kappa}{e^{\kappa z\cosh{t}}-1}\equiv J_{num}(v)+ \widetilde{J}(v) \,.
 \end{split} 
\end{equation}
The first term is the purely numerical one, the second one instead is easy to compute,
\begin{equation}
    z^2\frac{6}{\pi^2}\pi \int_0^{\infty}d\kappa\, \int_0^{\infty}dt\frac{\kappa}{e^{\kappa v\cosh{t}}-1}= 1\, , 
\end{equation}
while the last term involves the analytical part of the phase shift. Explicitly, 
\begin{equation}
     \widetilde{J}(z)=1+\frac{24}{\pi} z^{2}\int_0^{\infty}d\kappa\, \int_0^{\infty}dt\int_0^{\infty}du\frac{d}{d\kappa}\left( \cosh(u) \frac{\kappa^2 e^{2 \pi \kappa \cosh(u)}}{(e^{2 \pi \kappa \cosh(u)}-1)^2}\right)\frac{\kappa}{e^{\kappa z\cosh{t}}-1}\,  , 
\end{equation}
which we also have to evaluate numerically. We plot the results in Fig.\ref{fig:dTApp}. Let us consider the asymptotic behaviours. For $z\rightarrow\infty$ the leading contribution comes from the $l=0$ partial wave in the function $R_{l_{max}}(\kappa)$, which vanishes for small $\kappa$ as
\begin{equation}
    R_{l_{max}}(\kappa)\simeq -\frac{\pi}{2}\frac{1}{\log(\kappa b)}
\end{equation}
with $b\simeq 1.31$. The contribution to $J(z)$ is
\begin{equation}
    J(z)\simeq -\frac{3}{\pi^2}\int_0^{\infty}dv\, \int_0^{\infty}dt \frac{1}{\log(v b/z)}\frac{v}{e^{v\cosh{t}}-1}\simeq \frac{1}{2}\frac{1}{\log(z)}\, ,
\end{equation}
therefore for large $z$ the dominant contribution derives directly from the numerical part $J_{num}$. The slow decay is a direct consequence of the $l=0$ phase shift vanishing as $1/\log(\kappa)$. In the opposite limit $z\rightarrow 0$ we see that $J_{num}\rightarrow 0$ while $\widetilde{J}\rightarrow 1$ and the relevance of the two contributions gets exchanged. 

\section{The effective action in the stringy formalism}\label{app: stringy}

In this section we present the leading terms in the effective action describing the dynamics of the worldsheet Goldstone bosons with the bulk pions. The formalism has been introduced in section \ref{sec: 5} and consists on treating the zero modes of a given string soliton as collective coordinates depending on the worldsheet. 
For rotational invariant solutions, there are two normalizable zero modes associated to the broken translation symmetry, comprised in the complex scalar field $z_0$. The resulting effective action is naturally expanded in powers of the inverse cutoff $g$. 
Such an expansion becomes manifest by performing the following rescaling for the quantum fluctuations
\be\label{eq: rescaling}
z_0 \to \frac{1}{\sqrt{T_{cl}}}Z_0=\frac{g}{\sqrt{4\pi }}Z_0 \quad , \quad \delta \omega \to \frac{g}{\sqrt{2}}\delta \omega
\ee
such that the resulting kinetic terms are canonically normalized and the effective action takes the form
\be
S= g^{-2}S_{cl} + S^{(2)} + g S^{(3)} + \cO\left( g^2 \right) 
\ee
where the subscript indicates the order in fluctuations. Note that we are restricting this analysis to the unit charge string since, otherwise, the extended moduli would lead to additional worldsheet scalar fields. 

The leading term is of course the classical action of the soliton
\be
g^{-2}S_{cl}= \frac{2}{g^2} \int d^4x \frac{\partial_\mu \omega_{cl}\partial^{\mu}\overline\omega_{cl}}{\left(1+|\omega_{cl}|^2\right)^2}={\rm Vol}(\Sigma_w)T_{cl}
\ee
and the order $g^{-1}$ vanishes due to the fact that we are expanding around a solution to the equations of motion. 

The kinematics of the quantum fluctuations is governed by the quadratic terms which, by means of \eqref{eq: rescaling}, arise at order $g^0$ and read
\be
S^{(2)} = \int d^2\sigma \partial_\mu Z_0 \partial^\mu \bar Z_0 + S_\omega^{(2)} 
\ee
with
\bea
S^{(2)}_\omega =\int d^4x \frac{1}{(1+|\omega_{cl}|^2)^2}&\left[\partial_{\mu}\delta \omega\partial^\mu \delta\overline \omega-\frac{2\left( \partial_z \omega_{cl} \partial_{\bar z}\delta \overline \omega + {\rm c.c.}\right)\left(\omega_{cl}\delta\overline \omega+{\rm c.c.}\right)}{1+|\omega_{cl}|^2} \right. \\
& \qquad \left. -\frac{\partial_z \omega_{cl} \partial_{\bar z}\overline \omega_{cl} \left(2(1+|\omega_{cl}|^2)|\delta\omega|^2 -3(\omega_{cl}\delta\overline \omega +\overline \omega_{cl} \delta\omega)^2\right) }{(1+|\omega_{cl}|^2)^2}\right]
\eea
At this order there are potential terms mixing the Goldstone and the bulk pion fields. These are of the schematic form
\be 
-\int d^2\sigma d^2 z \frac{\partial_\sigma Z \partial_z \omega_{cl} \partial^\sigma\delta \overline \omega }{\left(1+|\omega_{cl}|^2\right)^2} +{\rm c.c.}=\int d^2\sigma d^2 z \frac{\partial^2_\sigma Z \partial_z \omega_{cl} \delta \overline \omega }{\left(1+|\omega_{cl}|^2\right)^2} +{\rm c.c.}
\ee 
However, these terms vanish due to the orthogonality of the bulk fluctuations and the zero modes (see section \ref{sec: 5}).

Finally, the leading interactions at order $g$ are given by the following cubic vertices
\be
S^{(3)}=S^{(3)}_{Zw} + S^{(3)}_{w}
\ee
where
\bea
S^{(3)}_{Zw}=\int d^4x\frac{1}{(1+|\omega_{cl}|^2)^2}&\left[ \frac{1}{\sqrt{2}2\pi }\partial_z \omega_{cl}\partial_\sigma Z_0\left( \partial^\sigma Z_0\partial_z\delta\overline \omega + \partial^\sigma \bar Z\partial_{\bar z}\delta\overline \omega \right) + {\rm c.c.} \right. \\
&\left. \,\, -\frac{1}{\sqrt{4\pi }} \partial_\sigma \delta \omega \left( \partial^\sigma Z\partial_z\delta\overline\omega + \partial^\sigma \bar Z\partial_{\bar z}\delta\overline \omega\right)  + {\rm c.c.} \right]
\eea
and
\be
S^{(3)}_\omega =-\int d^4x \frac{1}{(1+|\omega|^2)^4}\left[ \partial_z \omega_{cl}\partial_{\bar z}\delta \overline\omega \left((1+|\omega_{cl}|^2)|\delta \omega|^2 -3(\omega_{cl}\delta\overline\omega +\overline \omega_{cl} \delta \omega)^2 \right) + {\rm c.c.} \right]
\ee
At order $g^2$ there are various quartic vertices involving two Goldstone fields and two bulk pions as well as four bulk pions and we will not display these terms here.

The vertices depicted above determine the leading contribution to the to scattering cross sections for processes involving both NG bosons and bulk pions as external states. As recently noticed in \cite{Aharony:2024ctf}, these processes may hide several subtleties, and the ability to study them within the effective field theory has yet to be addressed. This is an interesting problem that we do not attempt to address in this work.

\bibliographystyle{ytphys}
\baselineskip=0.85\baselineskip
\bibliography{bib}

\end{document}